\documentclass[twocolumn]{aastex631}

\usepackage{amsmath}
\usepackage{amsfonts,amssymb}
\usepackage{txfonts}
\usepackage{gensymb}
\usepackage{natbib}

	\shorttitle{Spatial Spectrum of Solar Convection}
	\shortauthors{Getling \& Kosovichev}
	
\begin{document}	
	\title{Spatial Spectrum of Solar Convection from Helioseismic Data: Flow Scales and Time Variations}
	\correspondingauthor{A. V. Getling}
	\email{A.Getling@mail.ru}
	
\author{Alexander V. Getling}
\affil{Skobeltsyn Institute of Nuclear Physics, Lomonosov Moscow State
	University, Moscow, 119991 Russia}

\author{Alexander G. Kosovichev}
\affiliation{New Jersey Institute of Technology, NJ 07102, USA}

	\begin{abstract}
We analyze spectral properties of solar convection in the range of depths from 0 to 19~Mm  using subsurface flow maps obtained by the time--distance helioseismology analysis of solar-oscillation data from the Helioseismic and Magnetic Imager (HMI) onboard Solar Dynamics Observatory (SDO) from May 2010 to September 2020. The results reveal a rapid increase of the horizontal flow scales with the depth, from supergranulation to giant-cell scales, and support the evidence of large-scale convection, previously detected by tracking the motion of supergranular cells on the surface. Furthermore, the total power of convective flows correlates with the solar activity cycle. During the solar maximum, the total power decreases in shallow subsurface layers and increases in the deeper layers.
    \end{abstract}
	
	
	\section{Introduction} \label{sec:intro}
	
As is known, plasma flows in the solar convection zone form variously scaled structures resembling convection cells. The smallest of them, granules, have been known since the advent of high-resolution telescopic observations of the Sun \citep{Herschel1800}. \citet{Frenkiel1952} performed the first analysis of the turbulence spectrum of solar convection and, in addition to the primary maximum corresponding to granulation, found a secondary maximum at long wavelengths corresponding to 15~Mm. Based on Doppler measurements of horizontal velocities (away from the disk center), \citet{Hart_1954} discovered signs of a pattern of cells with sizes an order of magnitude larger and living much longer than granules. \citet{Leighton_etal_1962} described them in greater detail and designated them as supergranules. For a recent review of studies of supergranulation, see \citet{Rincon2018}. Further, \citet{November_etal_1981}, using Doppler measurements of vertical velocities, detected mesogranulation---a system of cells intermediate between granules and supergranules in their sizes. The existence of the largest velocity-field structures, giant cells, was suggested long ago by \citet{Simon_Weiss_1968}. At nearly the same time, \citet{Bumba_etal_1964} noted indications of the presence of giant structures observing magnetic fields. However, giant cells were qualified to be hypothetical for over three decades.

From the theoretical point of view, \citet{Glatzmaier_Gilman_II_1981} anticipated the existence of such cells based on their analysis of the onset of convection in a rotating spherical shell, performed in the anelastic approximation. They also found a tendency for meridionally elongated, banana-shaped convection cells to develop as the density stratification and rotation rate increase. Possibilities of banana-shaped cells have been repeatedly noted since the early 1970-s \citep{Busse_1970}; see also a survey by \citet{Busse_2002}. In particular, such cells were demonstrated in laboratory experiments \citep{Busse_Carrigan_1974}. Giant cellular structures were also found in global magnetohydrodynamic simulations of the solar-convection-zone dynamo action \citep{Ghizaru_etal_2010}. \citet{Featherstone_Hindman_2016} simulated convection in the anelastic approximation using a spectral technique. They {argued that ``the supergranular scale emerges due to a suppression of power on larger spatial scales owing to the presence of deep, rotationally constrained convection'' and that ``giant cells in the traditional sense do not exist''}.

The earliest direct Doppler observations of giant cells were done by \citet{Beck_1998}. Later, \citet{Hathaway_etal_2013} revealed them using supergranules as tracers of the material flow.

Lastly, \citet{Abramenko_etal_2012} reported the detection of mini-granules, whose sizes vary extremely widely. These features yet remain very poorly studied.
	
The multiscale structure of solar convection raises some questions that can be resolved only using the information on the velocity field in the subphotospheric convection zone. In particular, hydrodynamic considerations \citep[see, e.g.,][]{Shcheritsa_etal_2018} along with the fact that smaller-scale convection cells are advected at the photospheric surface by the flows in larger-scale cells and can be considered tracers of the large-scale velocity field \citep[see, e.g.,][]{Muller_etal_1992,Rieutord_etal_2001, GetlingBuchnev2010,Hathaway_2021} suggest that the convective velocity field represents a superposition of differently scaled flows \citep{GetlingBuchnev2010}, in contrast to the idea of the mixing-length theory that the flows have unique characteristic scales increasing with depth. The progress of helioseismological research has made it possible to trace the structure and evolution of the subsurface flow field over a large depth range and several years.

The flow scales are characterized by the spatial spectra of the velocity field. In particular, \citet{Hathaway_1987} applied a spherical-harmonic transform to the photospheric Doppler velocity signal and investigated the spatial scales of supergranules and giant cells. \citet{Hathaway_etal_2000,Hathaway_etal_2015} continued this line of research. They noted a broadband nature of the convection spectrum and employed a data-filtering technique to isolate the granular and the supergranular scale. \citet{Greer_etal_2015} employed ring-diagram techniques to study the strength and spatial scale of convective flows in the near-surface shear layer. In particular, they found that the peak of the horizontal velocity spectrum shifts with depth from higher to smaller values of the spherical-harmonic degree.

The spectral composition of the velocity spectrum at various depths below the photosphere has not yet received sufficient attention. Our aim is to study the spatiotemporal structure of subphotospheric convection using the horizontal-velocity maps of the \emph{Solar Dynamics Observatory (SDO)}, available from the Joint Science Operations Center (JSOC, \url{http://jsoc.stanford.edu/data/timed/}). In particular, we construct the spatial spectra of the horizontal-velocity fields obtained from the time--distance helioseismology pipeline and analyze the time variations of the spectra in the course of the solar activity cycle.

Attempts of investigating variations in the convection patterns over the activity cycle are not numerous \citep[see, in particular, references in][]{Roudier_Reardon_1998, Muller_etal_2018}. \citet{Lefebvre_etal_2008} found that the granulation evolves with height in the photosphere but does not exhibit considerable variations in the activity cycle. In addition, \citet{McIntosh_etal_2011} studied the variation of the supergranular length scale over multiple solar minima. \citet{Muller_etal_2018} detected no significant variations in the granulation scale with the activity cycle. \citet{Ballot_etal_2021} have shown that the density and the mean area of granules experience an approximately 2\% variation in the course of the solar cycle, the density of granules being greater and the area being smaller at the solar maximum.

In this Letter, we show that, in the multiscale velocity field traced at various levels in the
convection zone, the differently scaled flow components are superposed with one another. The scale of giant cells is present in the velocity spectrum along with the smaller, supergranular scales (at this stage of work, we filter out the granular and mesogranular components, which can introduce undesirable noise in our results). We also demonstrate variations of the integrated power of the velocity field in the course of the solar cycle.
	
\begin{figure*} 
	\centering
    \vspace{-2cm}
    \includegraphics[width=0.2674\textwidth]{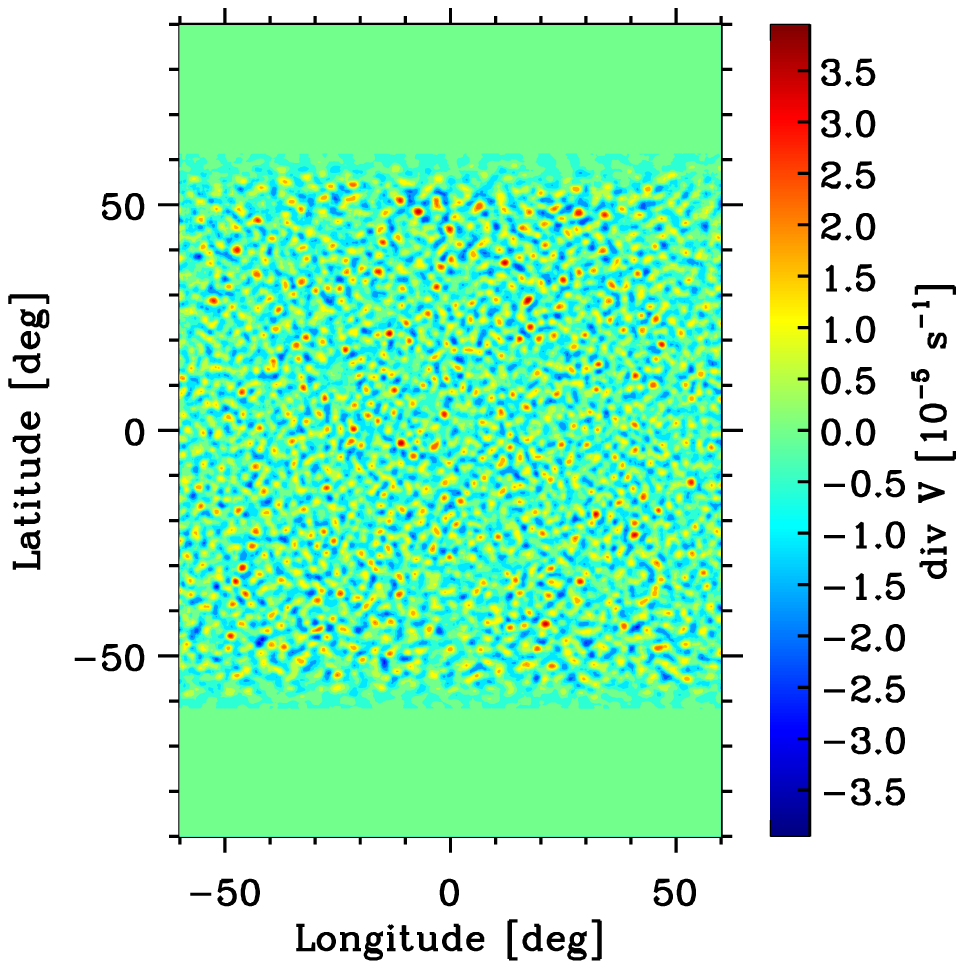}\quad
    \includegraphics[width=0.22\textwidth,bb=50 0 283 425,clip]{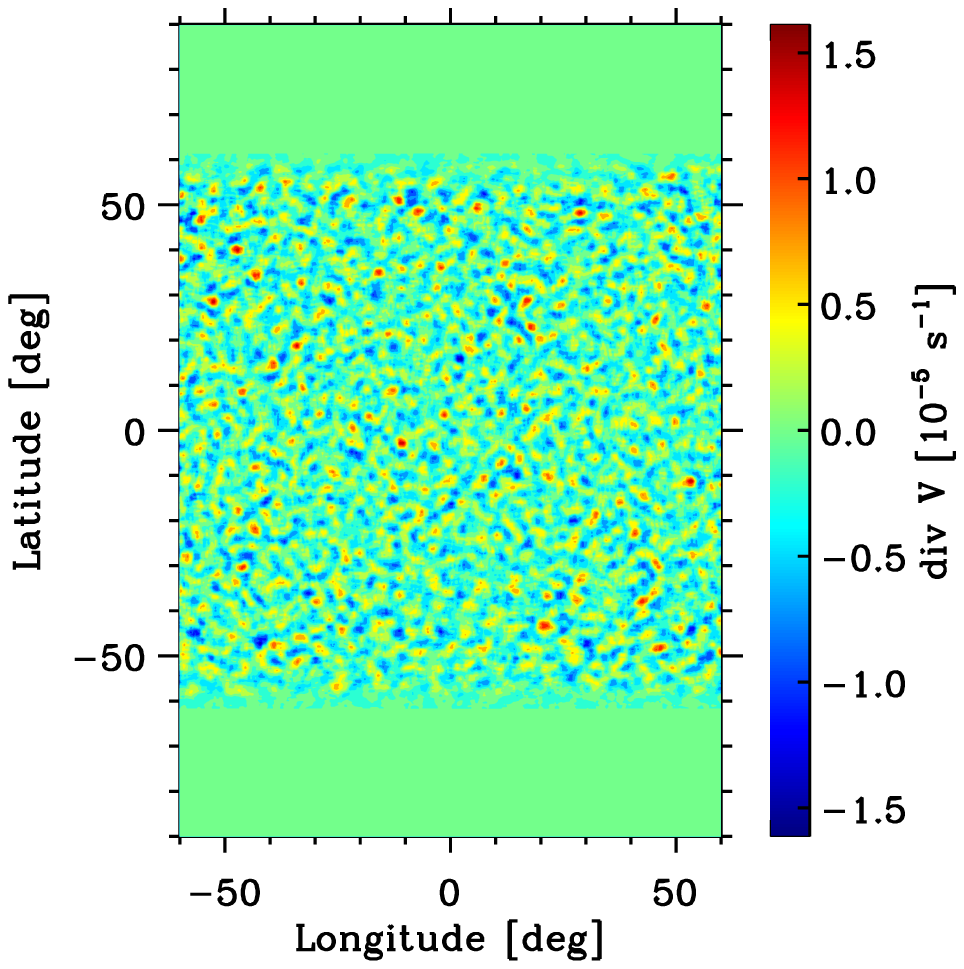}\quad
    \includegraphics[width=0.22\textwidth,bb=50 0 283 425,clip]{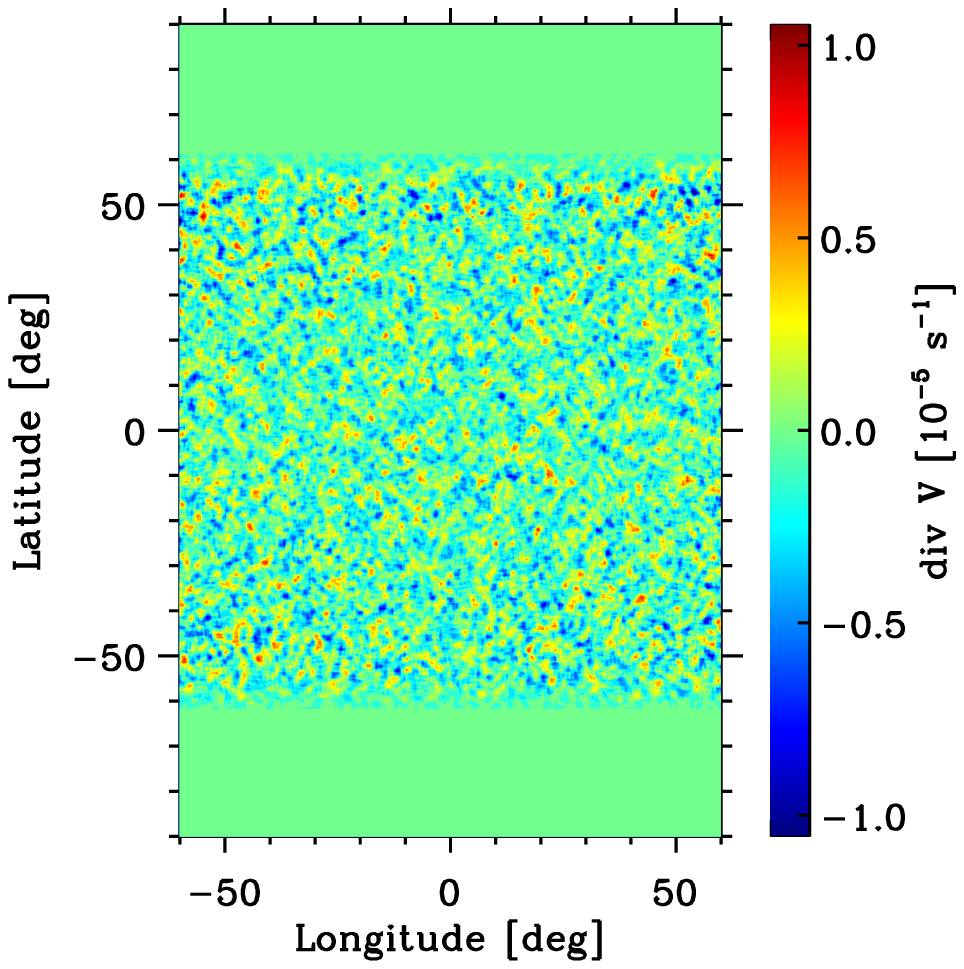}\quad
    \includegraphics[width=0.22\textwidth,bb=50 0 283 425,clip]{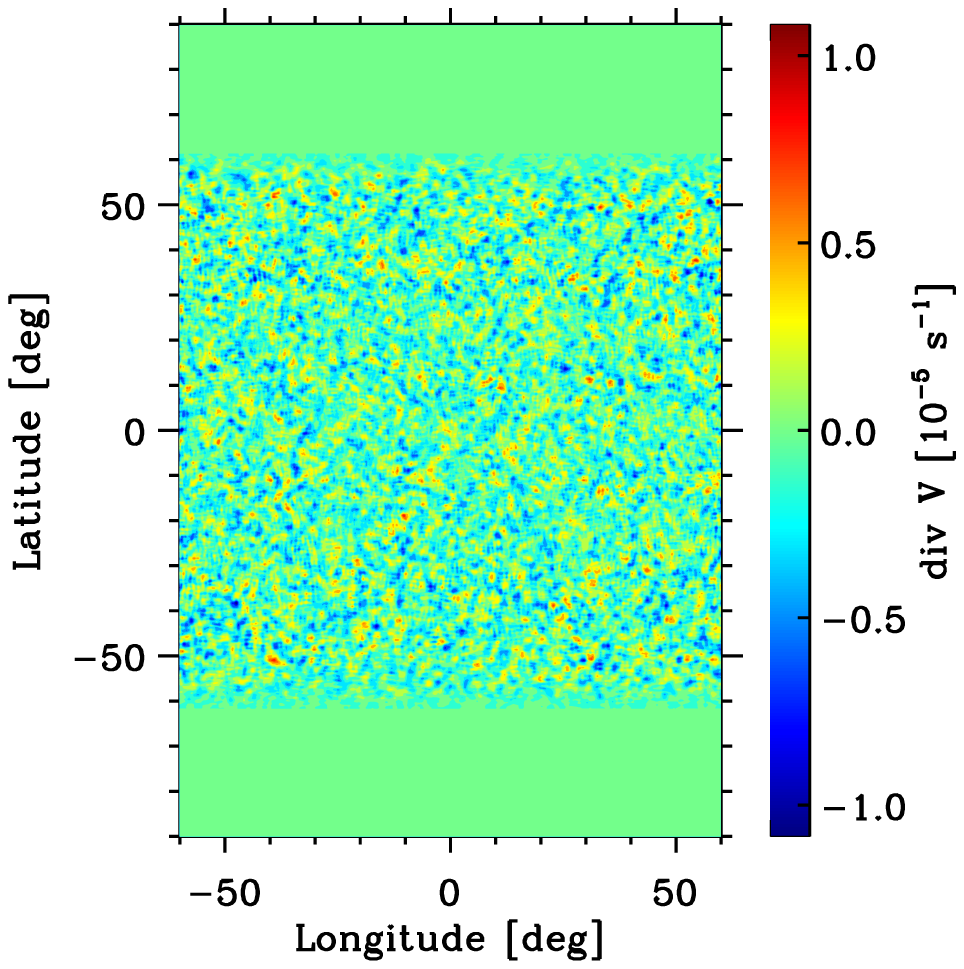}
    \caption{Sample maps of the divergence field at levels $d=0.5, 4.0, 6.0, 11.5$~Mm (from left to right) and the same time.
		\label{images}}
\end{figure*}

	\section{The Data and the Processing Techniques Used}\label{obs}
	
We use the subsurface flow maps for the central $120\degree \times 120\degree$ area of the visible surface of the Sun, routinely produced every 8 hours by the time--distance helioseismology pipeline \citep{Zhao_etal_2012,Couvidat_etal_2012} from the Helioseismic and Magnetic Imager (HMI) Dopplergrams \citep{Scherrer_2012,Schou_2012}. The flow maps used in our analysis are calculated for a grid of $1026\times 1026$ points spanning over 123\degree\ of heliographic latitude and Stonyhurst longitude with a spatial {sampling interval} of 0\degree.12 and a time cadence of 8~hours. We use data for the horizontal velocities at the following eight characteristic levels below the photosphere (the corresponding depth ranges for which the inversions were done are parenthesized): $d=0.5$ (0--1)~Mm, 2.0 (1--3)~Mm, 4.0 (3--5)~Mm, 6.0 (5--7)~Mm, 8.5 (7--10)~Mm, 11.5 (10--13)~Mm, 15.0 (13--17)~Mm, 19.0 (17--21)~Mm. The travel-time measurements are described by \citet{Couvidat_etal_2012}. The travel-time inversion procedure employed in the HMI pipeline uses Born-approximation sensitivity kernels and provides a good localization of the averaging kernels at the target depth. However, the {vertical} width of the averaging kernels increases with depth, from $\sim 2$~Mm near the surface to $\sim 10$~Mm at the bottom layer \citep[][{Figure 10}]{Couvidat_etal_2005}. The horizontal width of the averaging kernels also increases with depth, from $\sim 16$~Mm to $\sim 40$~Mm {\citep[][Figure 12]{Couvidat_etal_2005}}. Therefore, the flow maps represent the velocities convolved with the averaging kernels, and this should be taken into account in the interpretation of the presented results. {To assess the possible effect of the averaging-kernel variation with depth, we present an illustrative example in Subsection~\ref{scales}.}
		
	\subsection{Spectral Representations}
	
For convenience, we analyze the scalar fields of the horizontal-velocity divergence rather than the velocity vector $\mathbf V$. On a sphere of radius $r$, the divergence of the vector $\mathbf V=\{V_\theta, V_\varphi\}$ is
	\begin{equation}
		f(\theta,\varphi) = \mathop{\mathrm{div}}\mathbf{V}(\theta,\varphi) = \frac {1}{r\sin \theta}\frac {\partial}{\partial \theta}(V_\theta \sin \theta) + \frac {1}{r\sin \theta}\frac {\partial}{\partial \varphi}V_\varphi,
		\label{div}
	\end{equation}
where $\theta$ and $\varphi$ are the polar and azimuthal angles and $r$ is a meaningless constant divisor. We smooth the velocity field with a 17.5-Mm window to remove the smaller-scale components (the short-wavelength noise) and represent the field of the divergence as a polynomial expansion
	\begin{equation}
		f(\theta,\varphi) = \sum_{l=0}^{l_{\max}}\sum_{m=-l}^{l}A_{lm}Y_{l}^{m}(\theta,\varphi)
		\label{series}
	\end{equation}
in spherical harmonics of angular degree $l$ and azimuthal order $m$:
	\begin{gather}
		Y_{l}^{m} = \sqrt {\frac{(2l+1)}{4\pi}\frac{(l-m)!}{(l+m)!}} P_l^m (\cos \theta) \mathrm {e}^{\mathrm{i}{m\varphi}},
		\label{spherharm}
	\end{gather}
	$$l=0,...,l_{\max},\quad m=0,...,l$$
(where $P_l^m$ are the associate Legendre polynomials and $l_{\max}$  is the upper spectral boundary, which we assume to be equal to 200). The spectral coefficients (amplitudes of harmonics) can be determined in a standard way by the equations
	\begin{gather}
		A_{lm}=\frac{1}{4\pi}\int\limits_0^\pi\!\!\int\limits_0^{2\pi} f(\theta,\varphi)\,P_l^{m} (\cos \theta)\mathrm {e}^{\mathrm{-i}{m\varphi}} \sin \theta \ \mathrm d\varphi\, \mathrm d\theta.
		\label{coefts}
	\end{gather}
{Along with the spatial 17.5-Mm-window smoothing, we apply a running-averaging procedure with a 45-day window to our data---specifically, to the spectra $p_{lm}=|A_{lm}|^2$ and to the power functions $p_l$ and $p^\Sigma_l$ [see Equations~(\ref{powerph}) and (\ref{powertot}) below].}
	
Our source data do not cover the whole spherical surface. For this reason, we cut a longitudinal 120\degree\ band out of each flow map and complement it with the same data shifted by 120\degree\ and 240\degree\ to fill the complete longitudinal angle. {The resultant spectra thus contain  nonzero harmonics only with $m$ multiple of 3; we interpolate them to all missing $m$ values and smooth the spectra with a two-point window for better visual perceptibility.}
	
Since our source data are restricted to a latitudinal range of $\pm 61\degree.5$, we have to investigate the effect of the ``empty'' polar caps on the spectrum. We apply our spectral analysis to a sample model velocity field obtained by G.~Guerrero and A.M.~Stejko using numerical simulations (2021, private communication) and compare the spectra obtained with and without artificially introduced zero velocity in the polar caps $-90\degree < \varphi < -61.5\degree$ and $61.5\degree < \varphi < 90\degree$. In addition, we introduce a latitudinal tapering of the flow fields to reduce possible spurious effects due to the Gibbs phenomenon. In other words, we multiply the divergence fields by a window function, which smoothes the sharp drop of velocities at latitudes of $\pm 61.5\degree$---the boundaries of the ``empty'' polar caps. The analysis of the simulations shows that the ``empty'' polar caps only result in a moderate narrowing of the spectral $l$-band. The latitudinal tapering also results in minimal changes in the flow spectrum.
	
\begin{figure*} 
	\centering
    \includegraphics[width=0.33\textwidth,bb=50 0 800 650,clip] {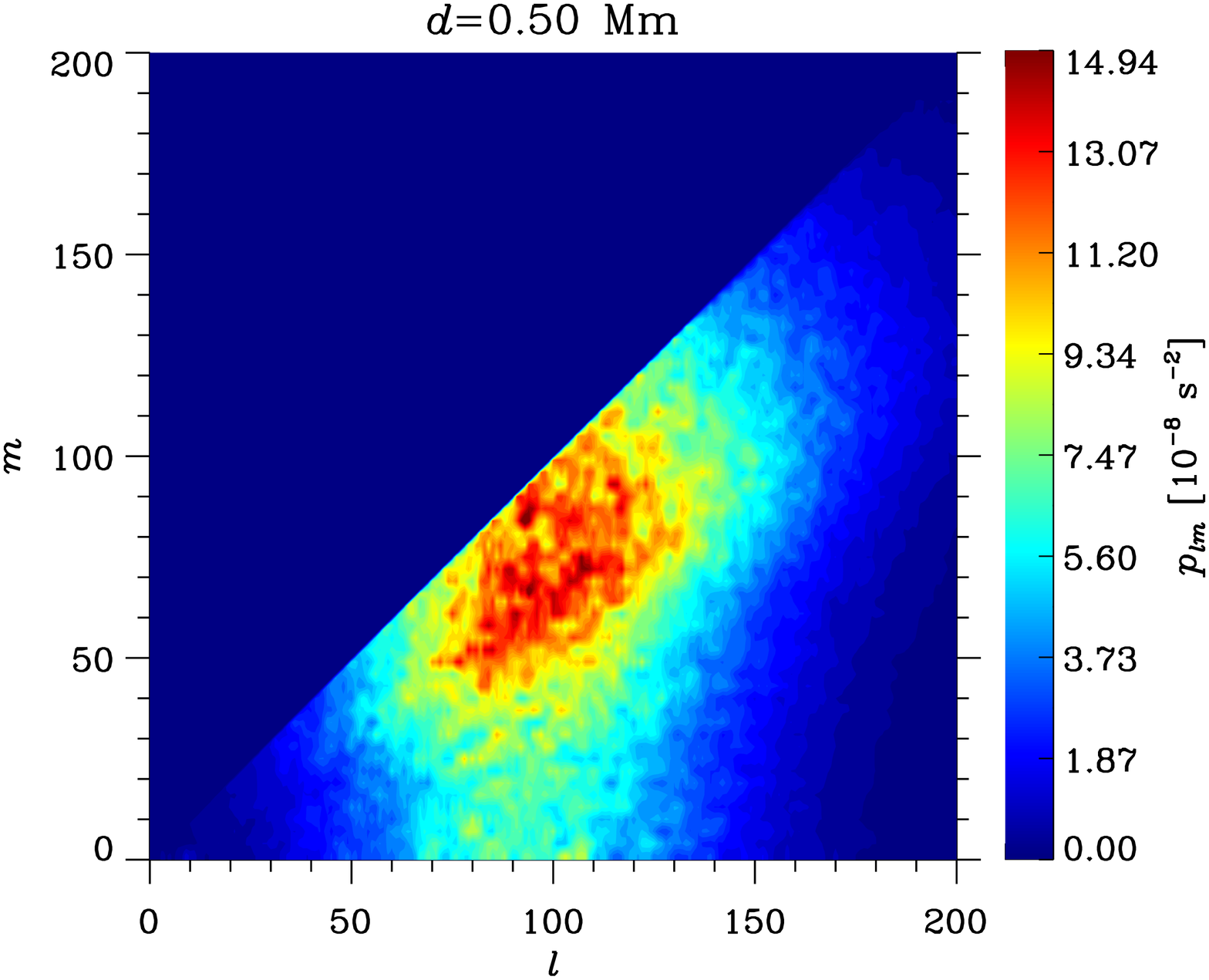}
    \includegraphics[width=0.33\textwidth,bb=50 0 800 650,clip] {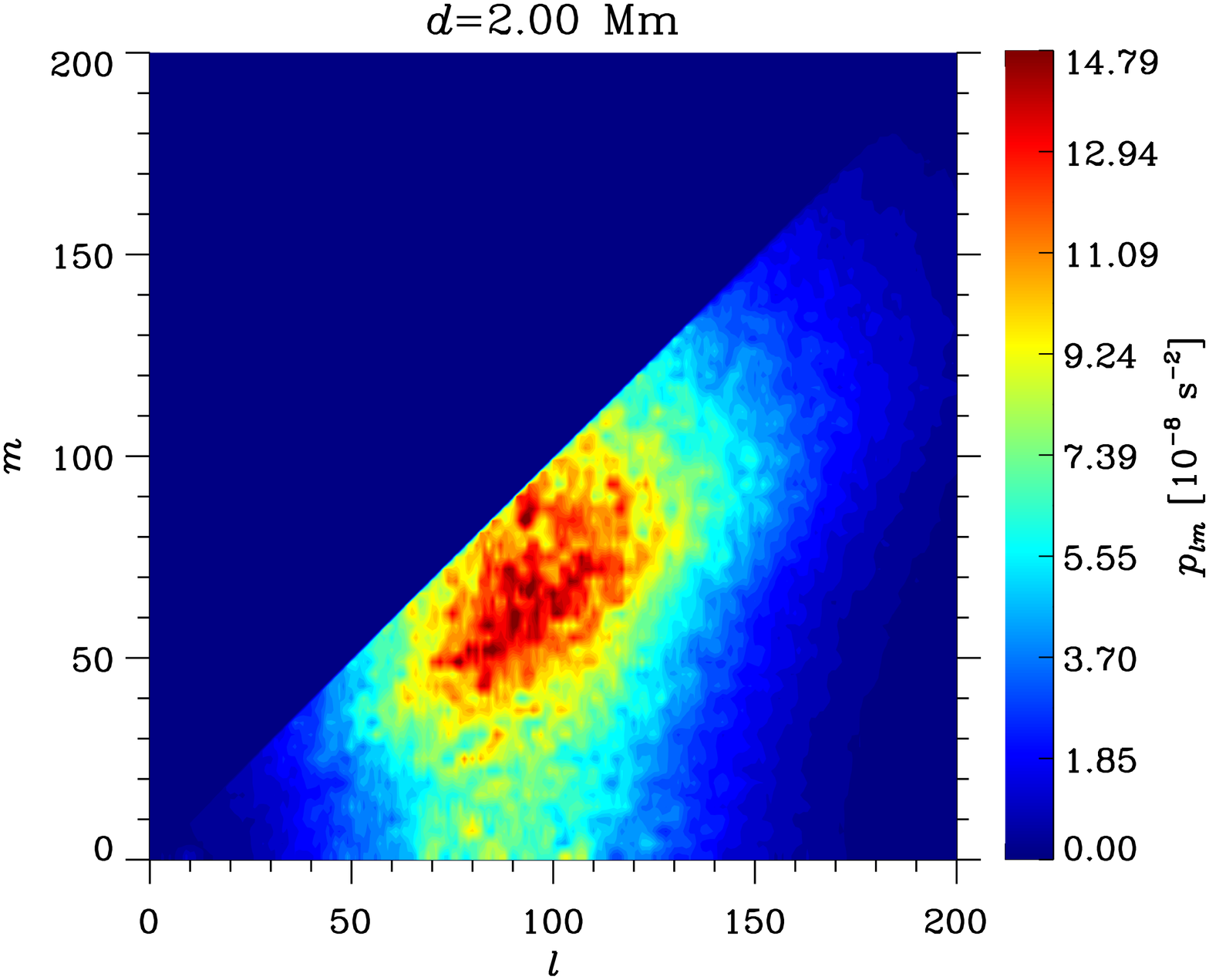}
    \includegraphics[width=0.33\textwidth,bb=50 0 800 650,clip] {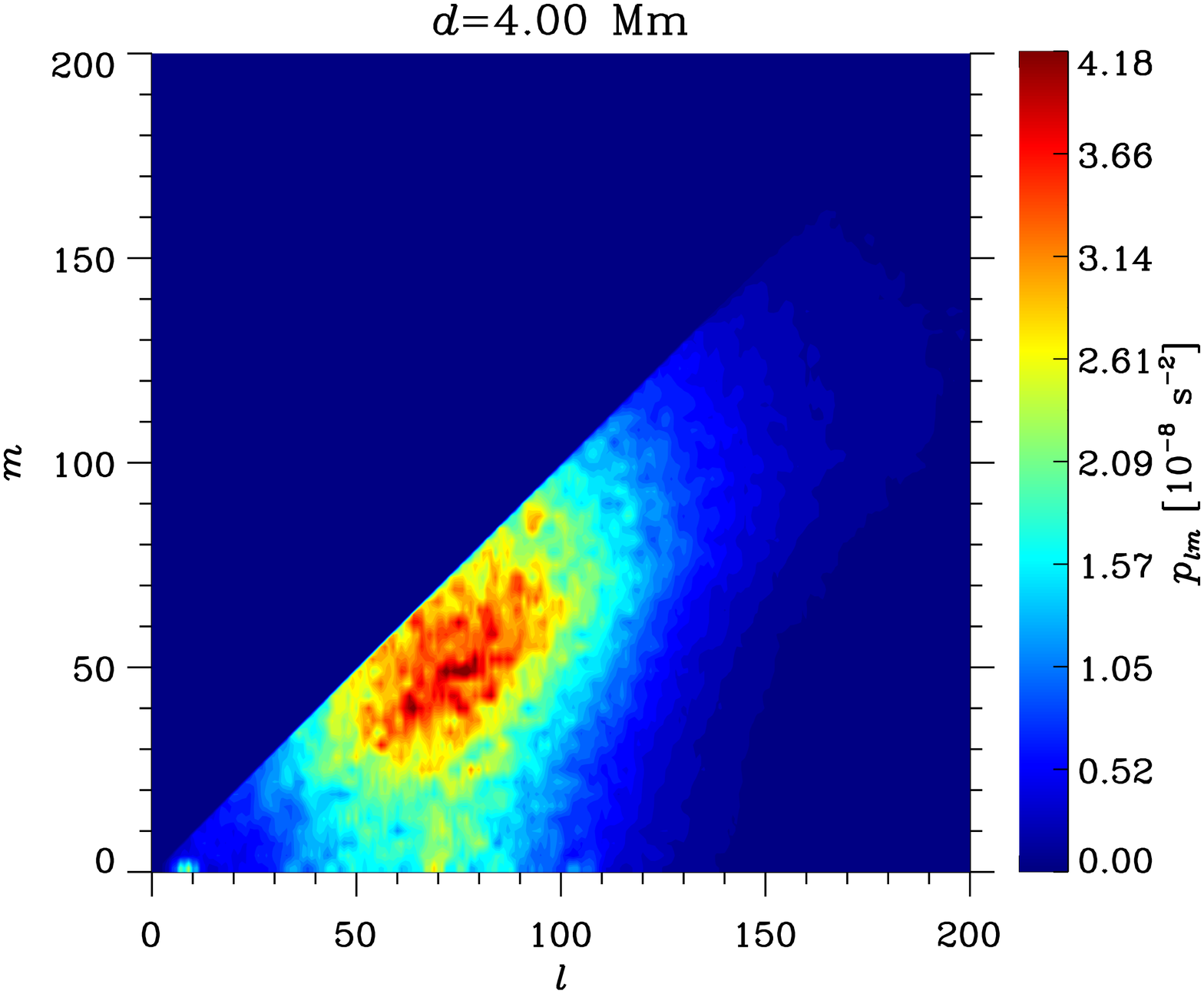}
    \includegraphics[width=0.33\textwidth,bb=50 0 800 650,clip] {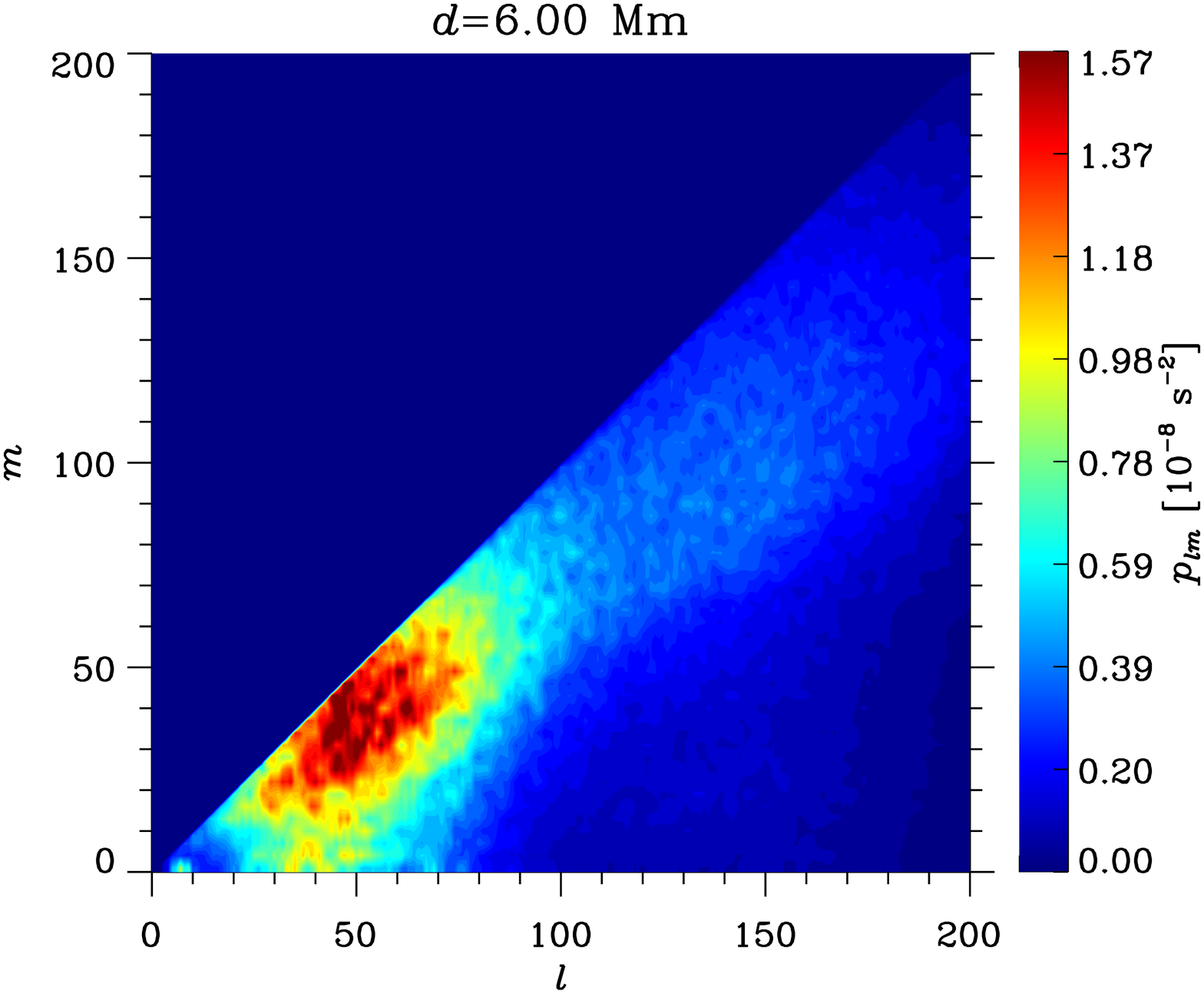}
    \includegraphics[width=0.33\textwidth,bb=50 0 800 650,clip] {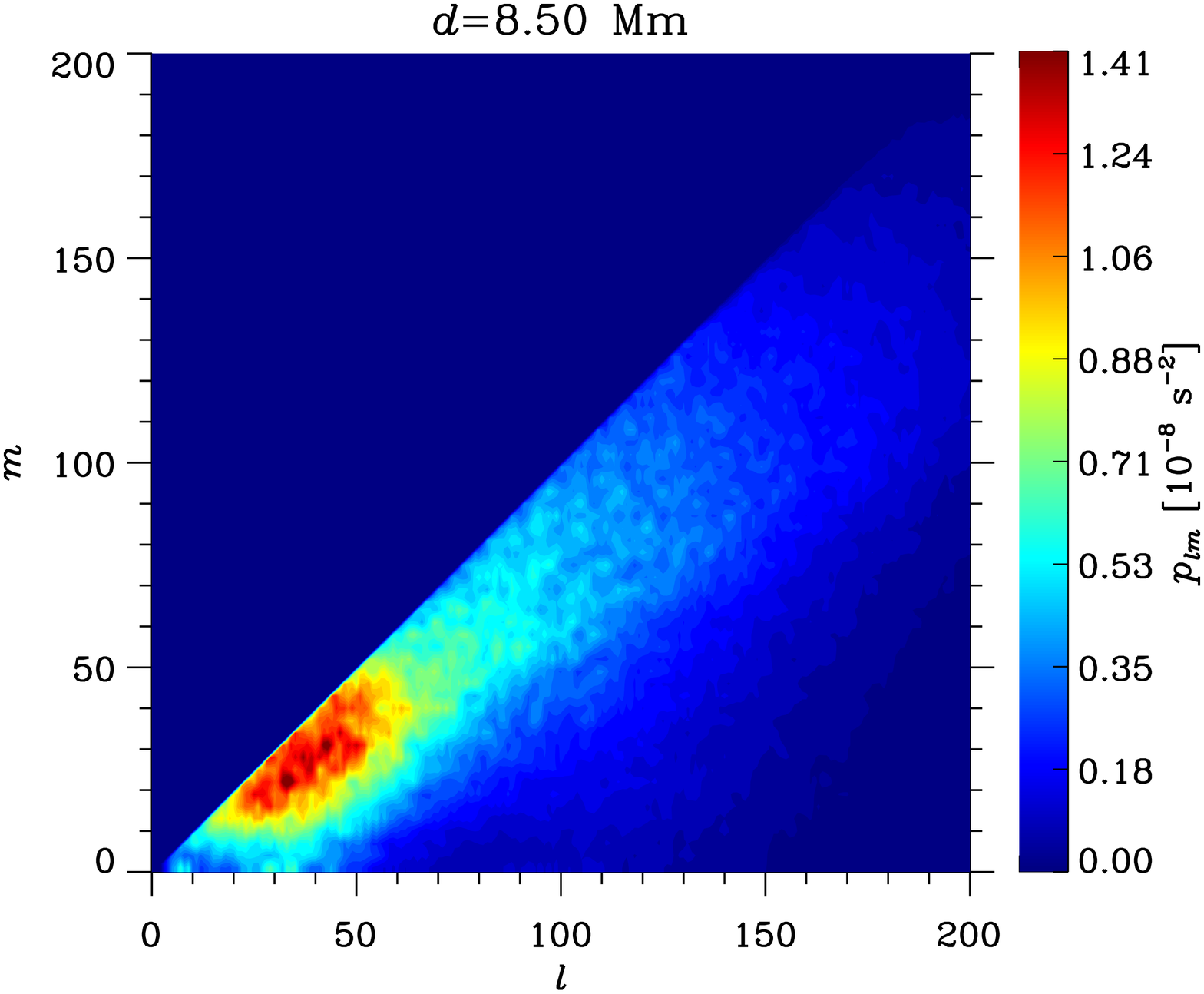}
    \includegraphics[width=0.33\textwidth,bb=50 0 800 650,clip] {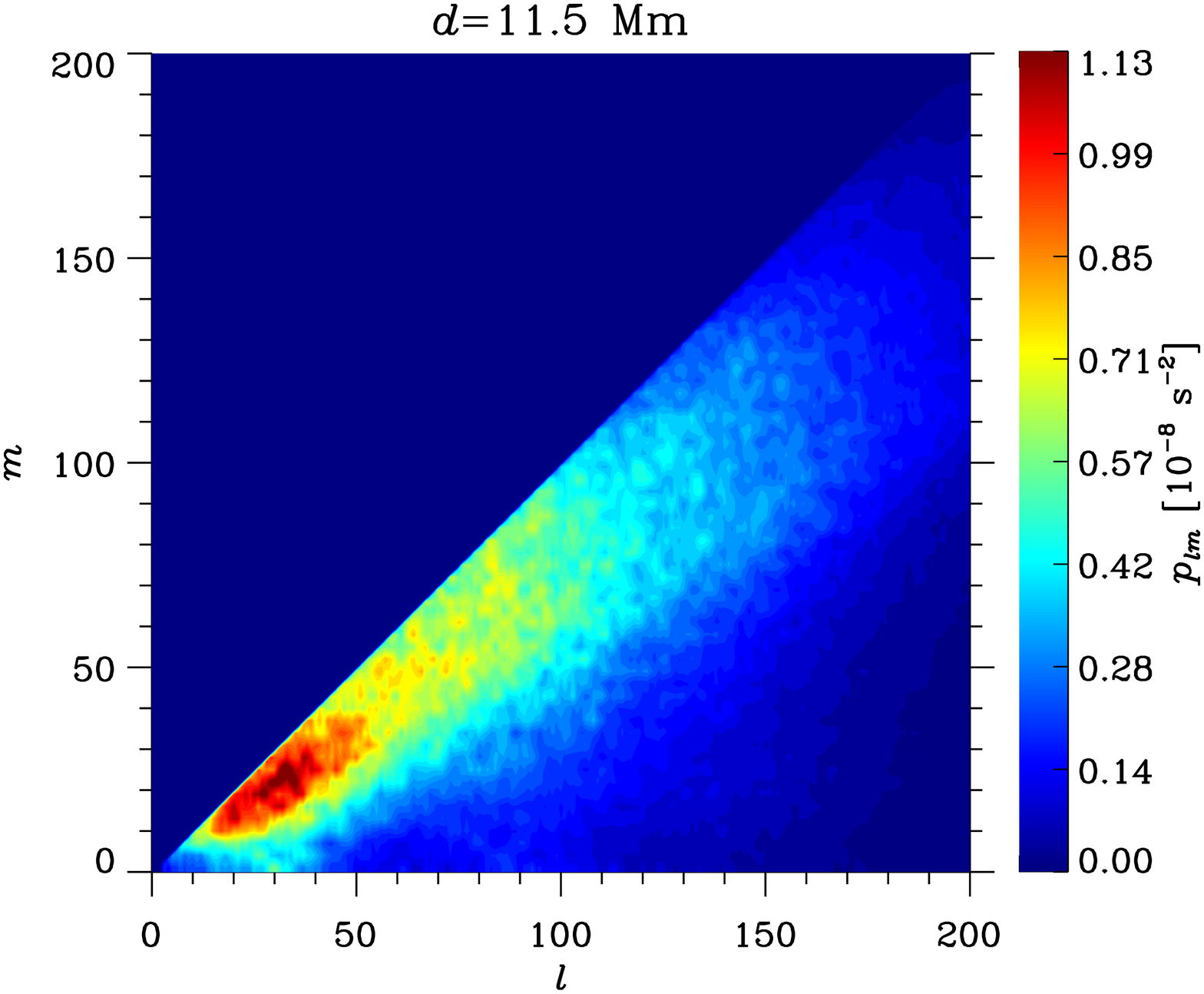}
    \includegraphics[width=0.33\textwidth,bb=50 0 800 650,clip] {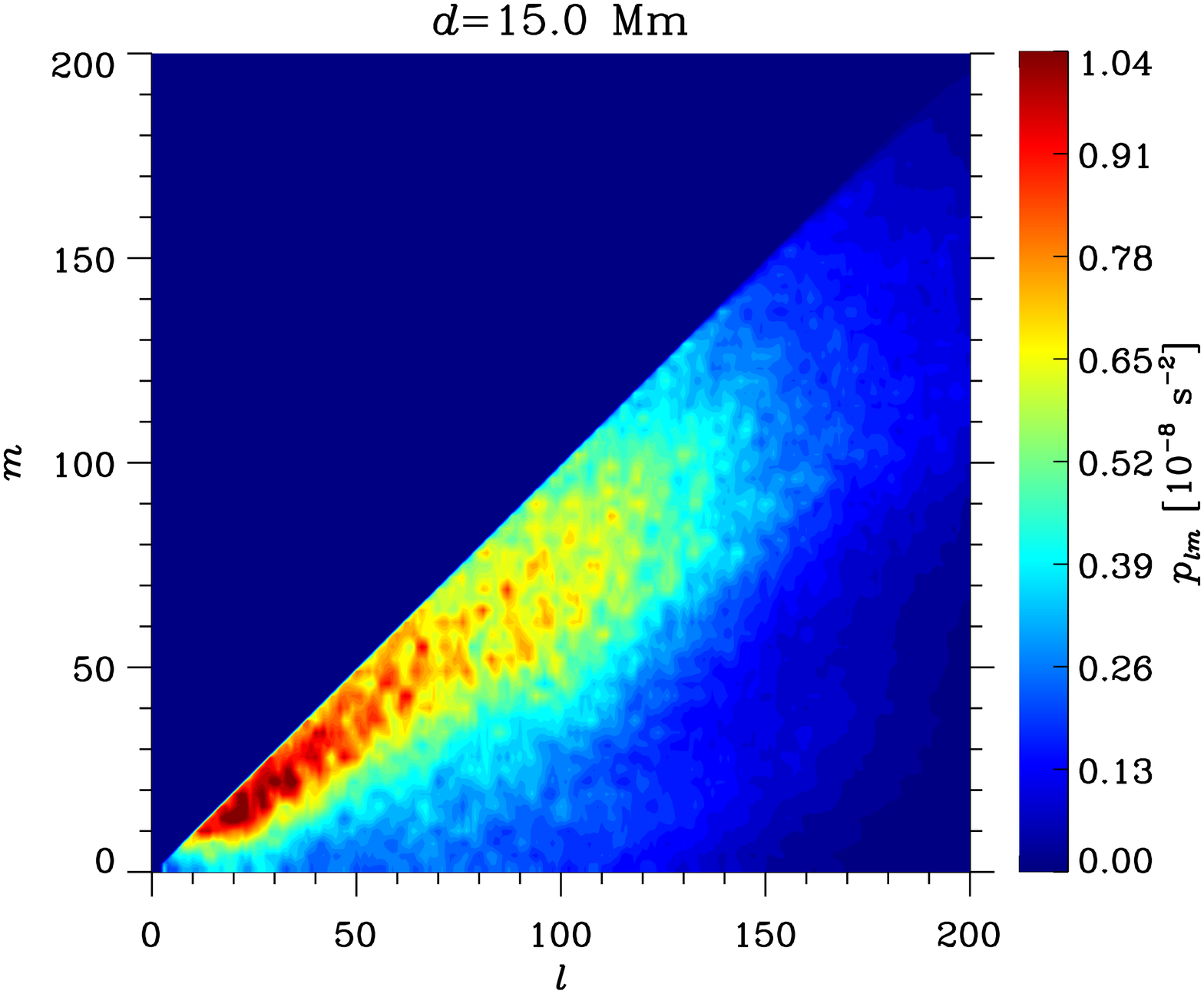}
    \includegraphics[width=0.33\textwidth,bb=50 0 800 650,clip] {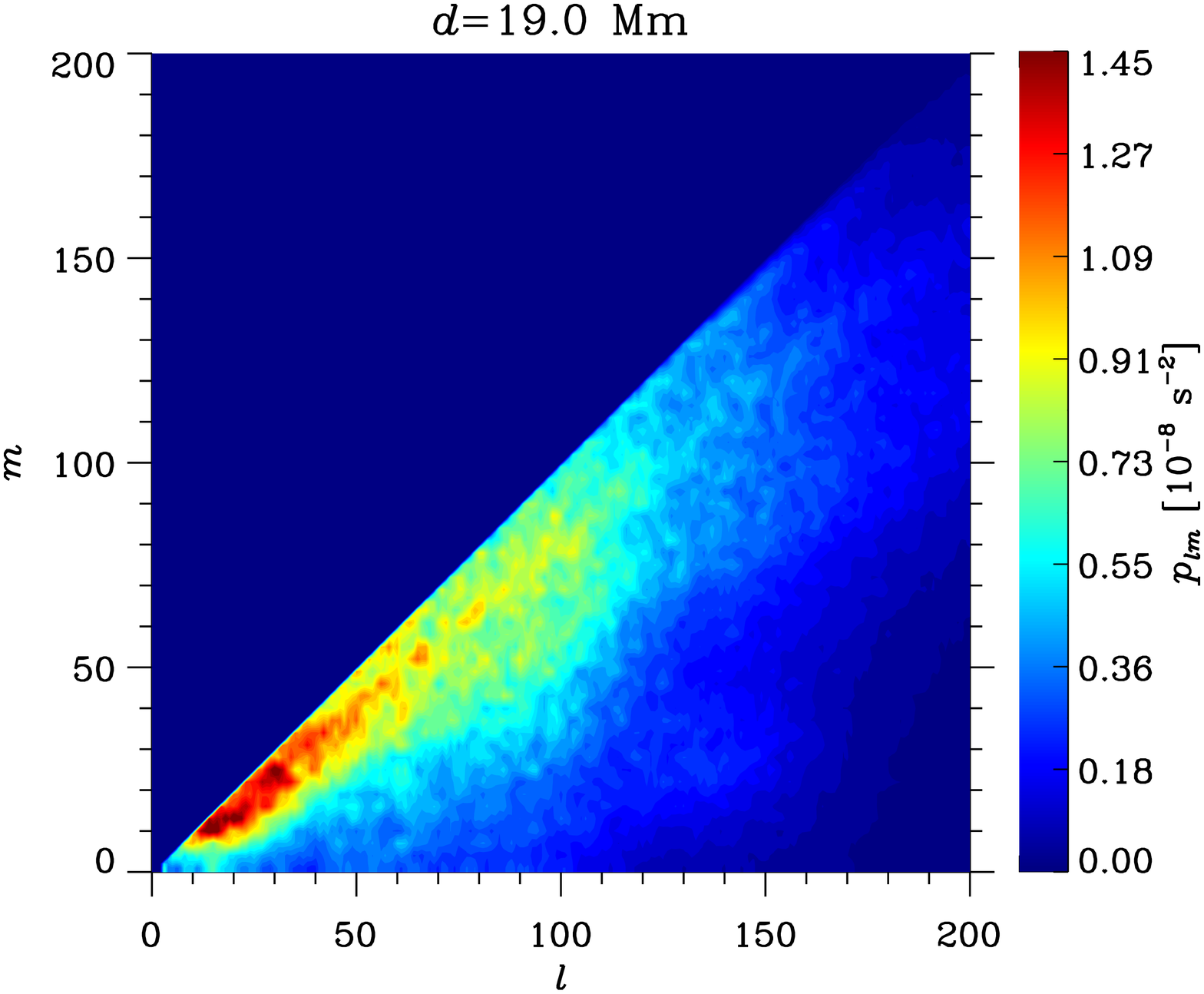}\\
\caption{Depth variation of a sample power spectrum, $p_{lm}$, for 2011 March 22--May 7 after applying 45-day running averaging. The depth values are indicated at the top of each panel.
		\label{spectra}}
\end{figure*}

\begin{figure*} 
\centering
\includegraphics[width=0.33 \textwidth,bb=20 0 435 226,clip] {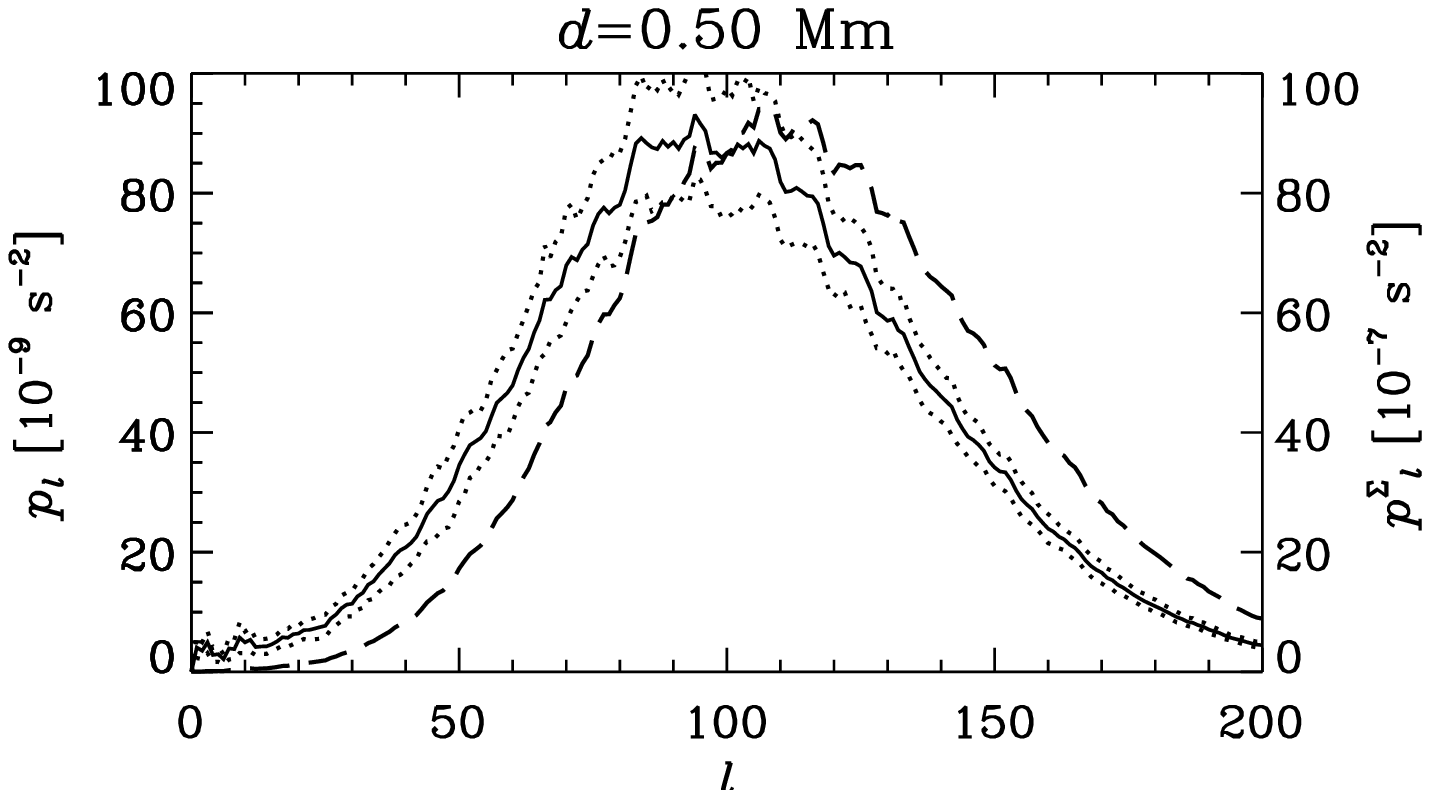}
\includegraphics[width=0.33\textwidth,bb=20 0 435 226,clip] {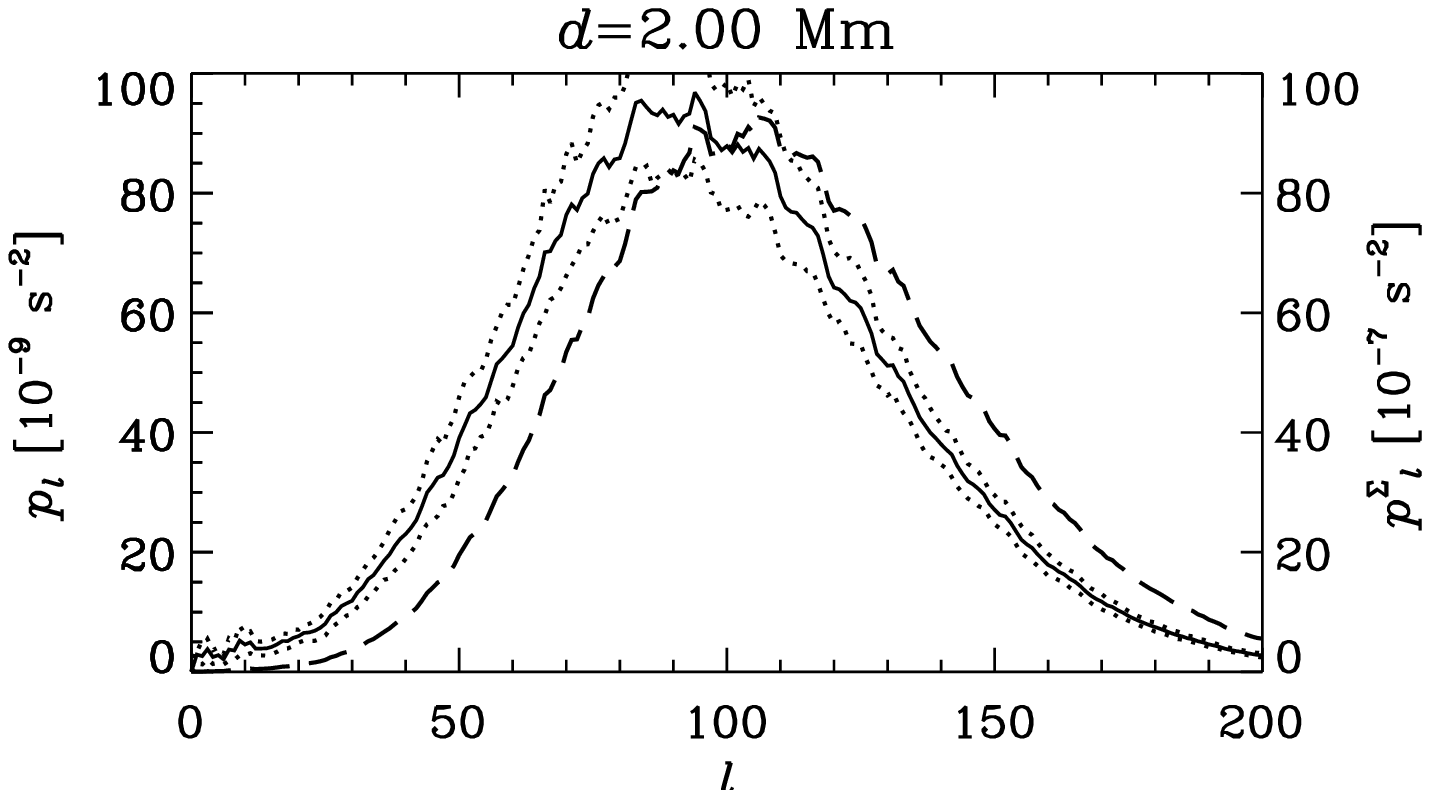}
\includegraphics[width=0.33\textwidth,bb=20 0 435 226,clip] {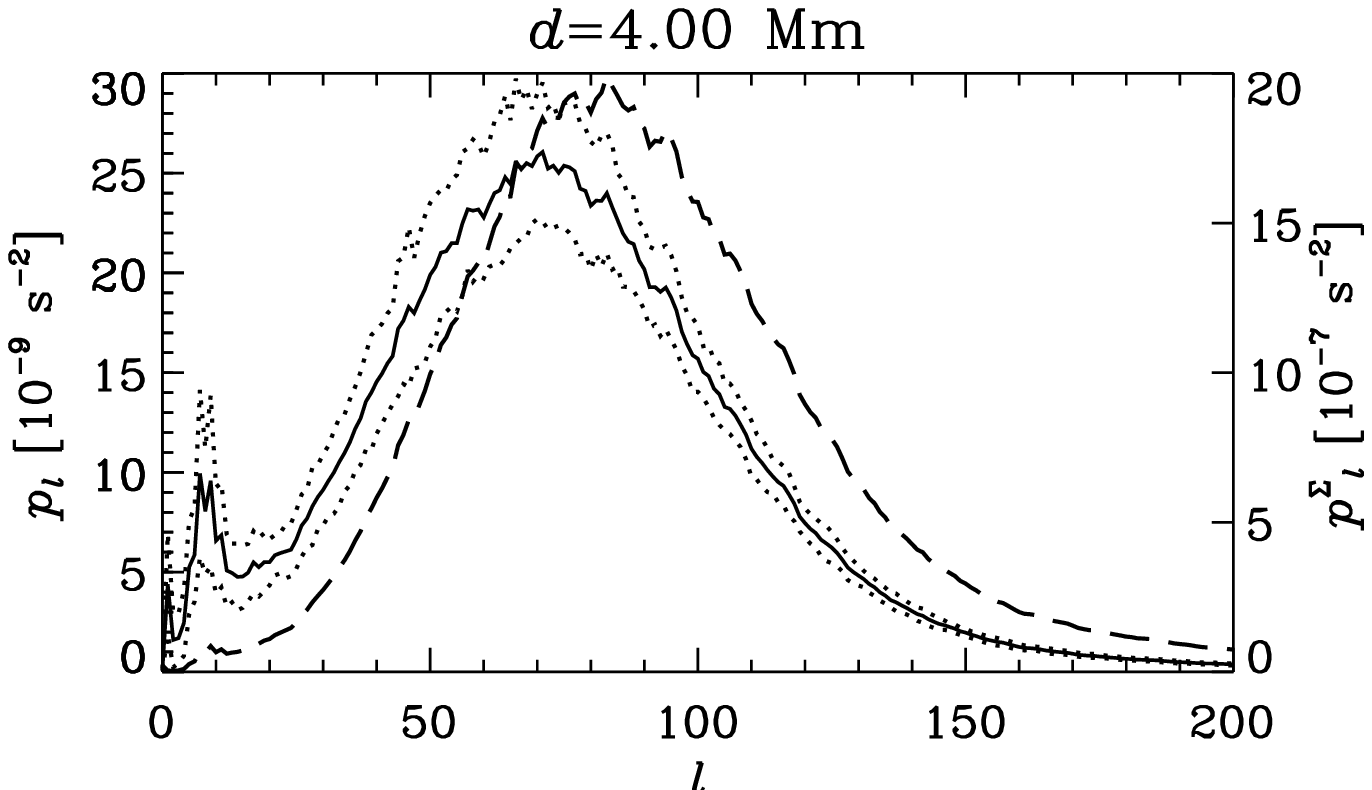}\\[6pt]
\includegraphics[width=0.33\textwidth,bb=20 0 435 226,clip] {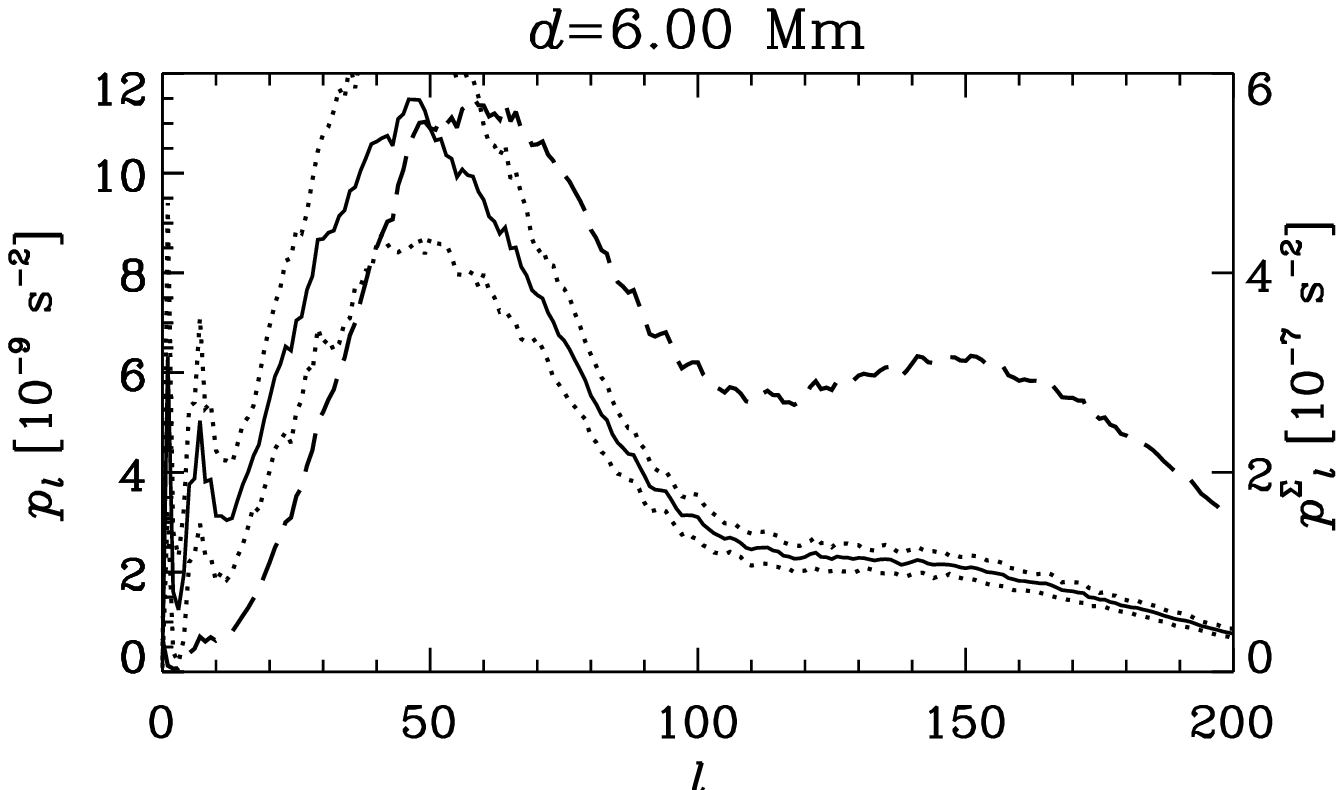}
\includegraphics[width=0.33\textwidth,bb=20 0 435 226,clip] {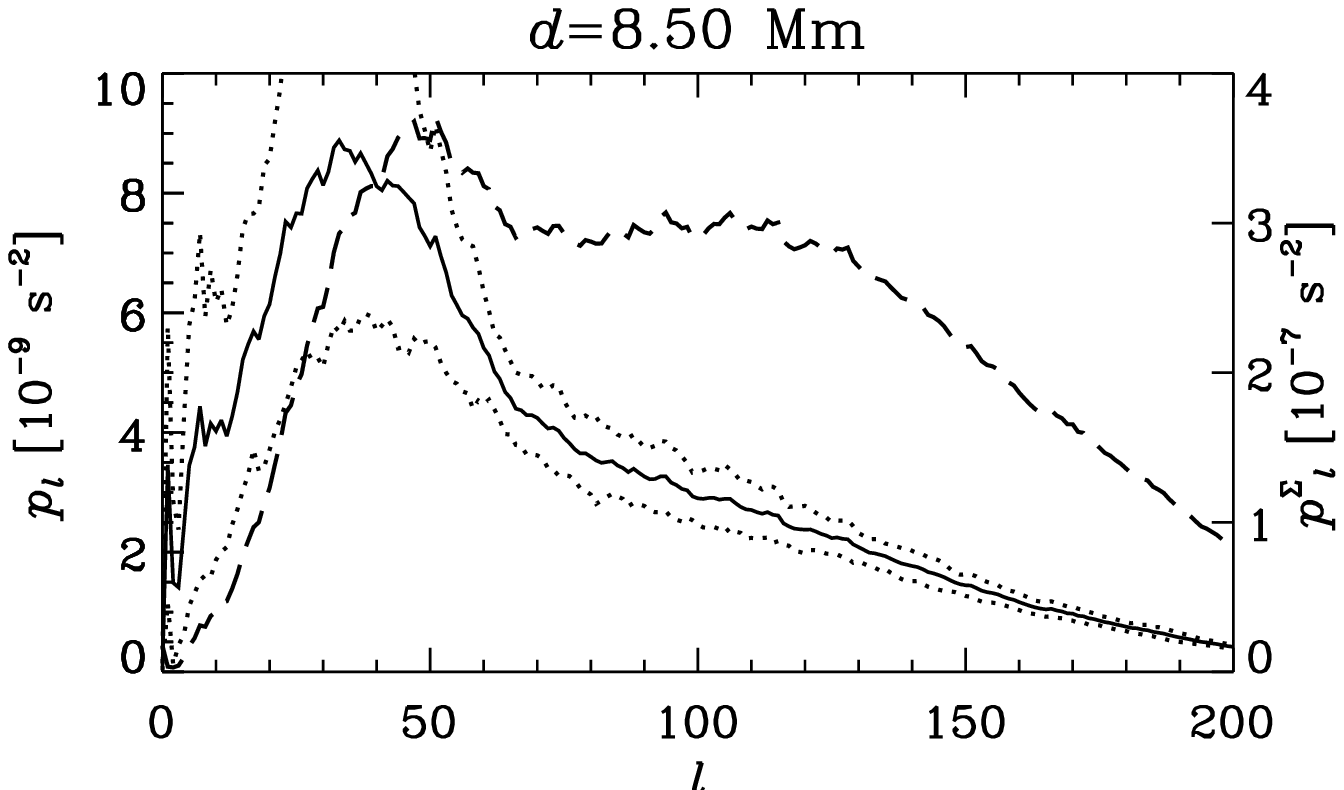}
\includegraphics[width=0.33\textwidth,bb=20 0 435 226,clip] {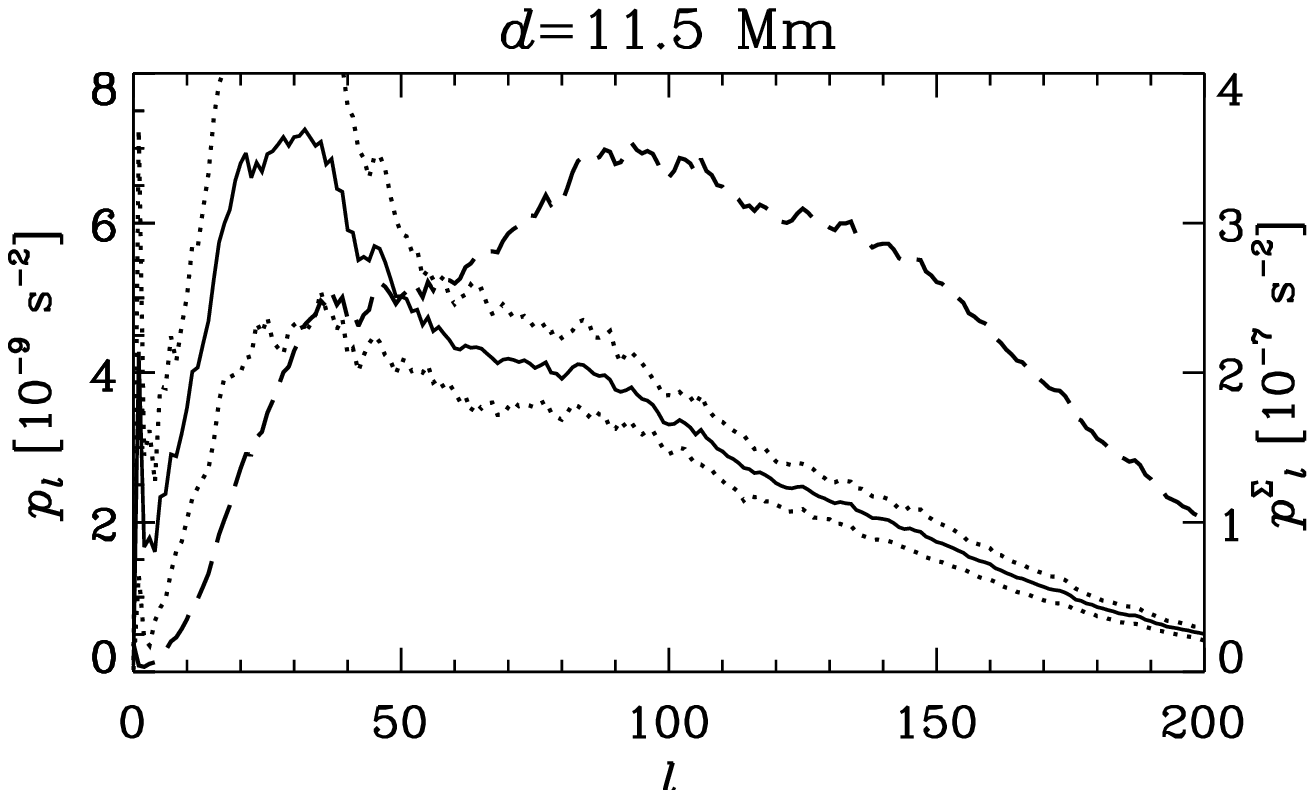}\\[6pt]
\includegraphics[width=0.33\textwidth,bb=20 0 435 226,clip] {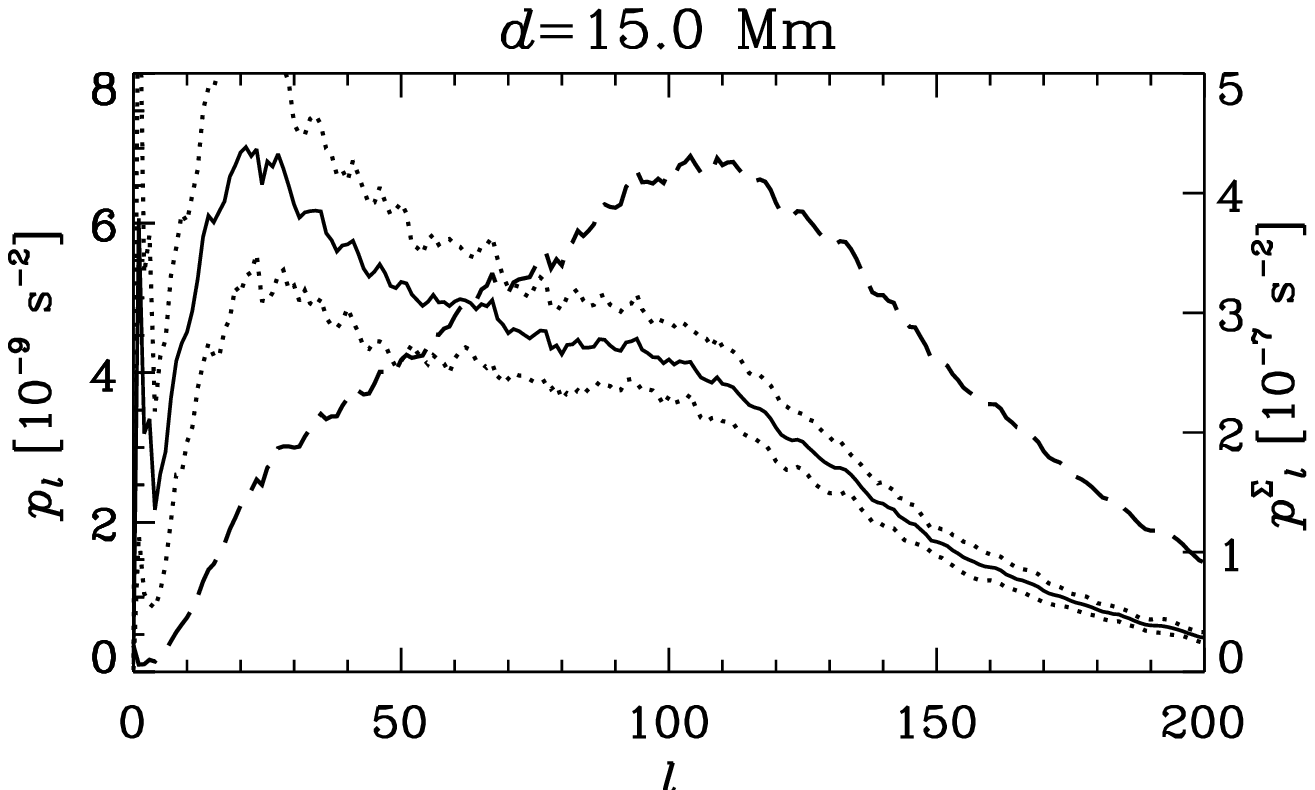}
\includegraphics[width=0.33\textwidth,bb=20 0 435 226,clip] {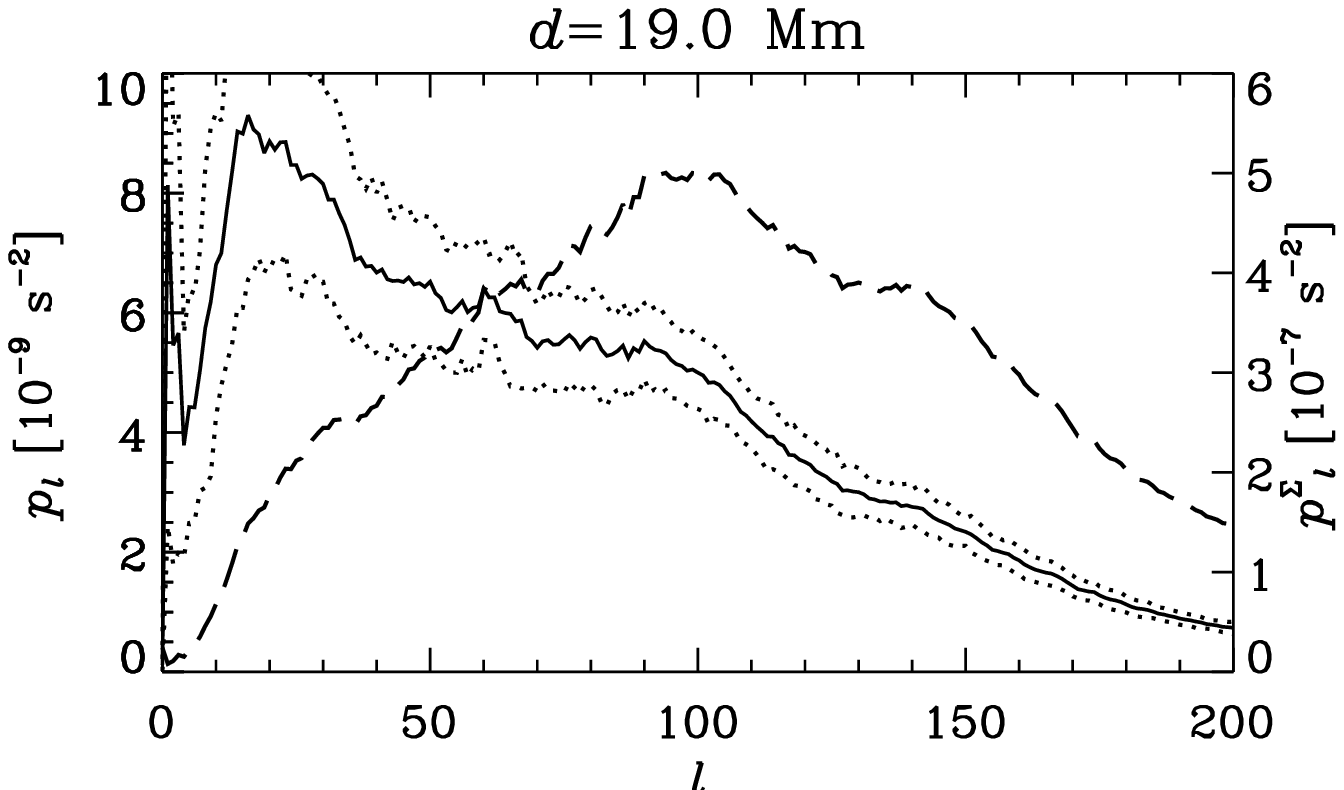}
\includegraphics[width=0.33\textwidth,bb=15 0 430 226,clip]{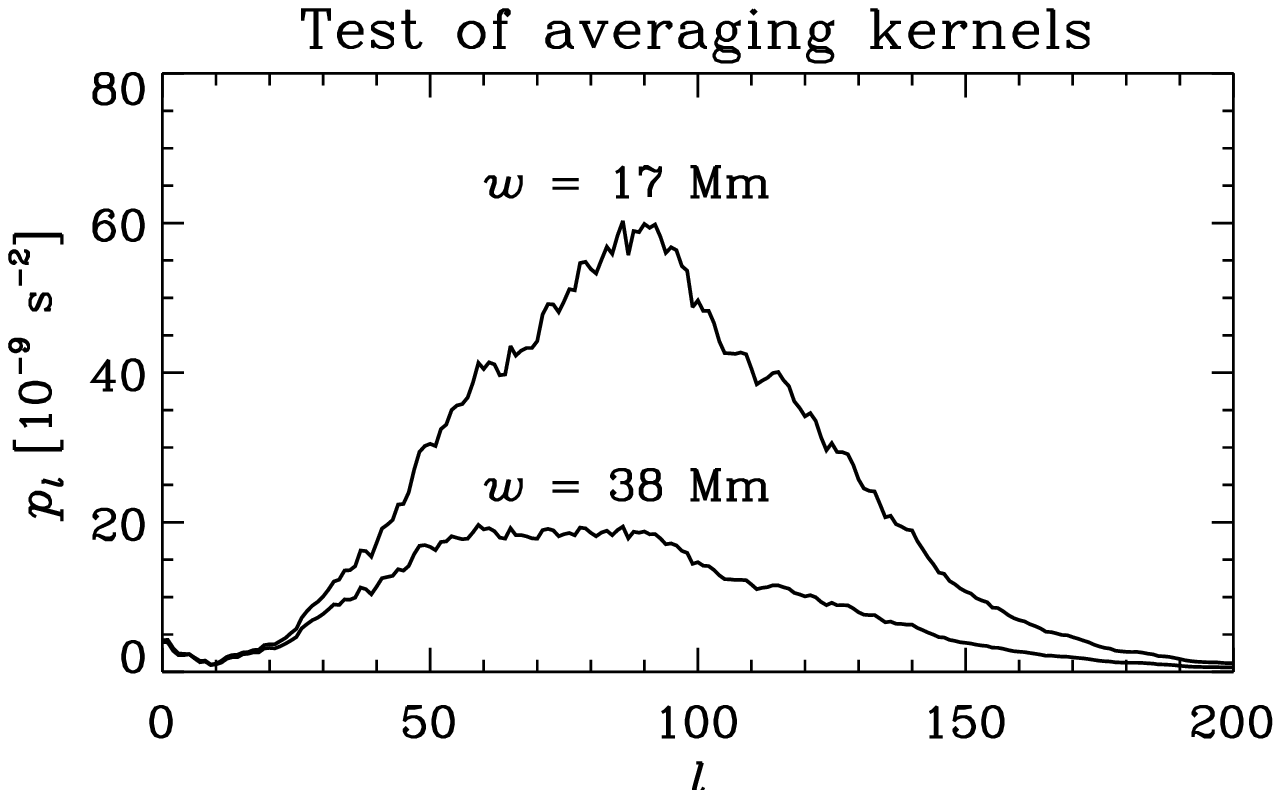}\\
\caption{Solid curves (in all panels but the last one): $l$-variation of the power $p_l$ based on the above-displayed spectra for different depths (Figure~\ref{spectra}); dotted curves indicate the standard deviation of $p_l$ from its running average; dashed curves: the total power, $p^\Sigma_l$. The depth values are indicated at the top of each panel. The bottom right panel shows the $p_l$ curves obtained for a velocity-divergence field at $d=0.5$~Mm averaged with the Gaussians corresponding to the helioseismic averaging kernels of two different widths,~$w$.
	\label{m-averages}}
\end{figure*}

\begin{figure*} 
	\centering
    \includegraphics[width=0.49\textwidth]{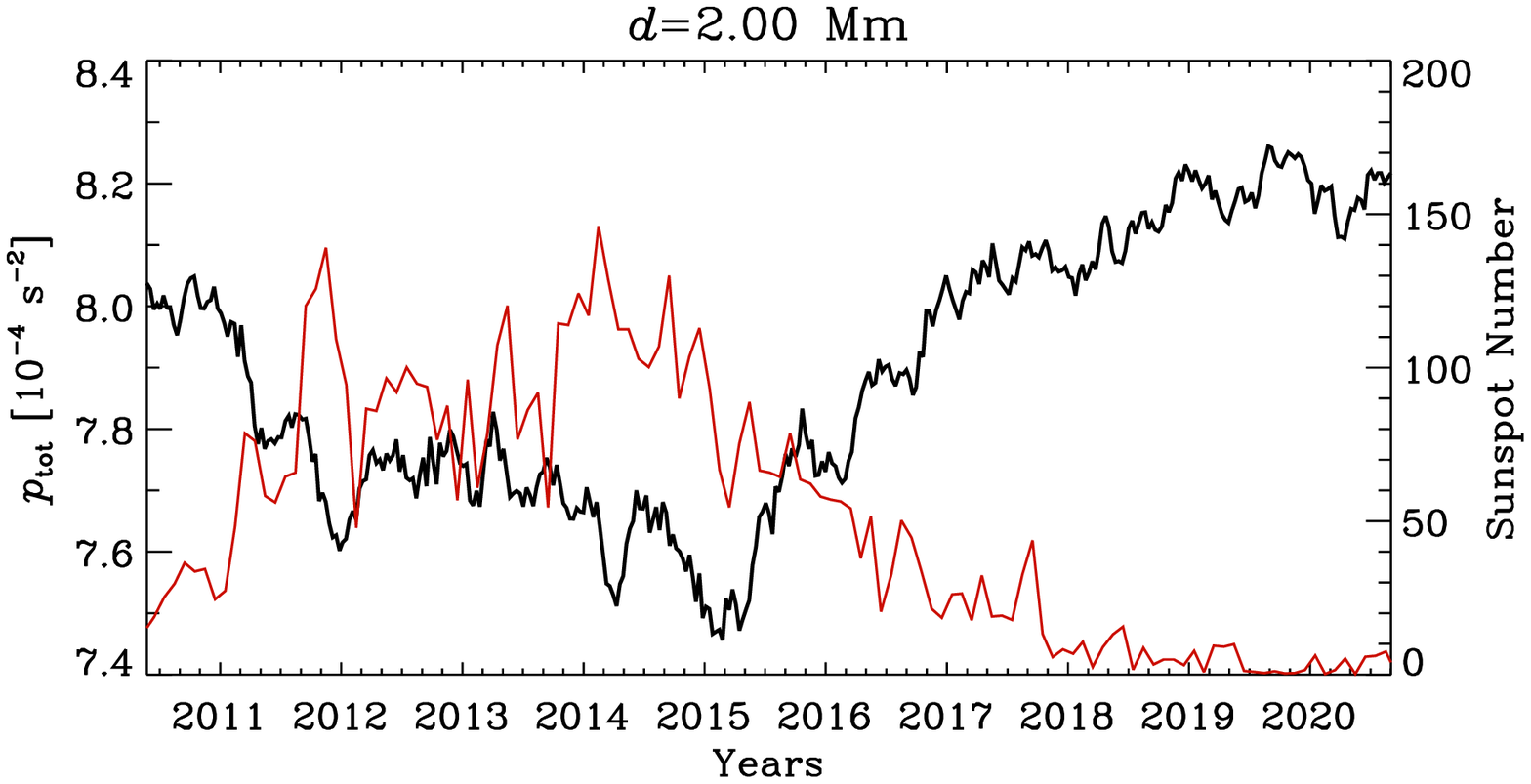}
    \includegraphics[width=0.49\textwidth]{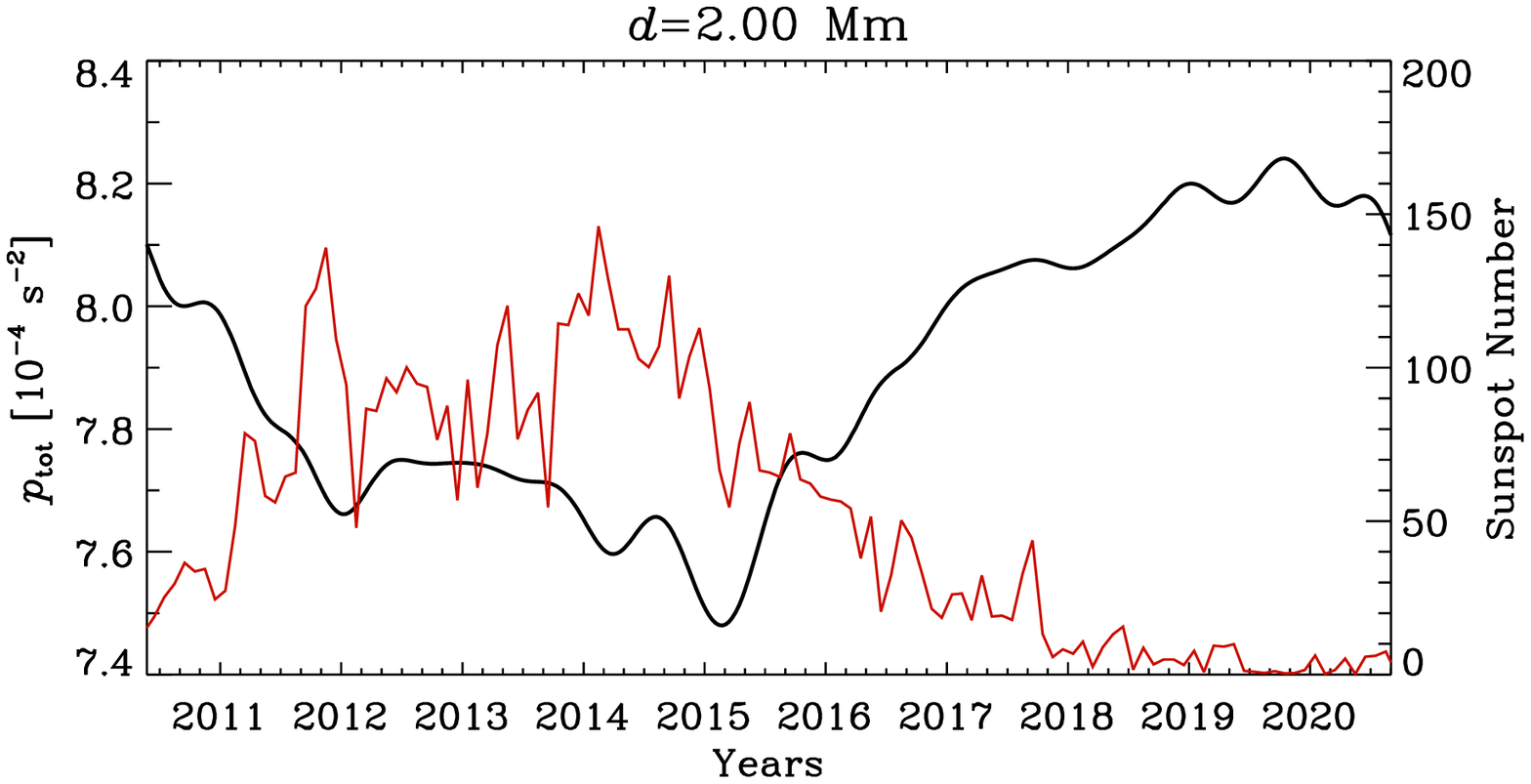}
    \includegraphics[width=0.49\textwidth]{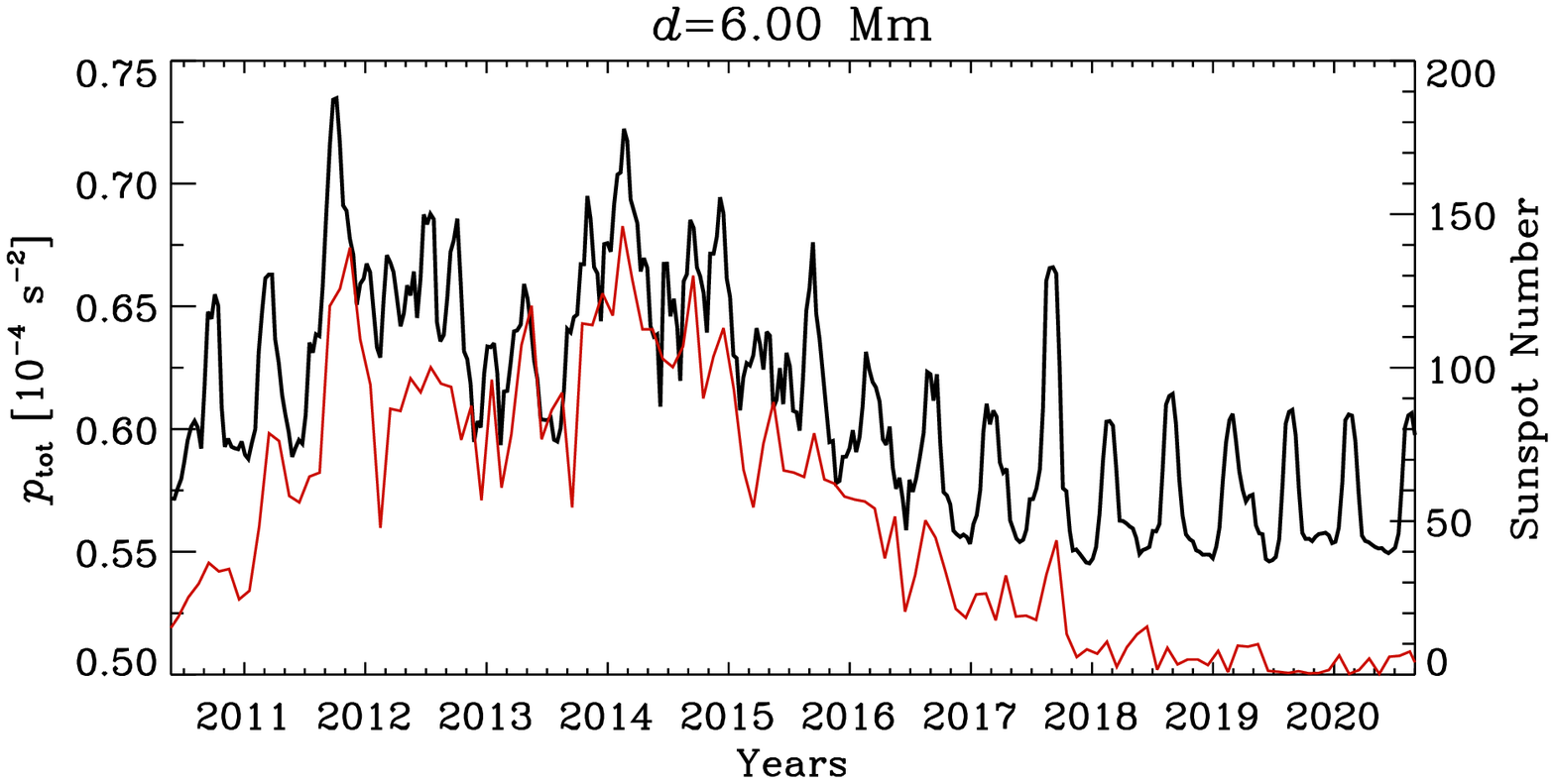}
    \includegraphics[width=0.49\textwidth]{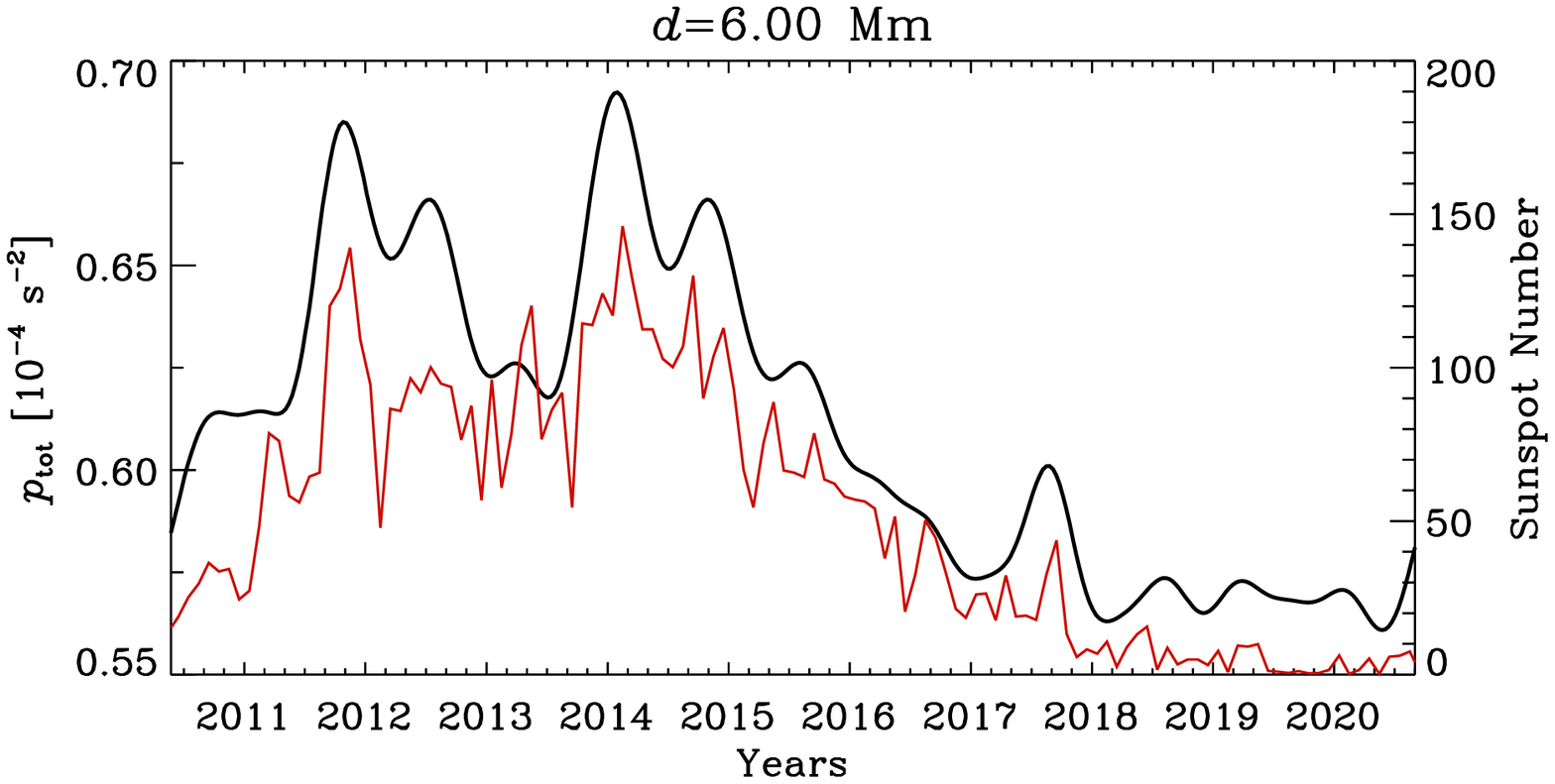}
    \includegraphics[width=0.49\textwidth]{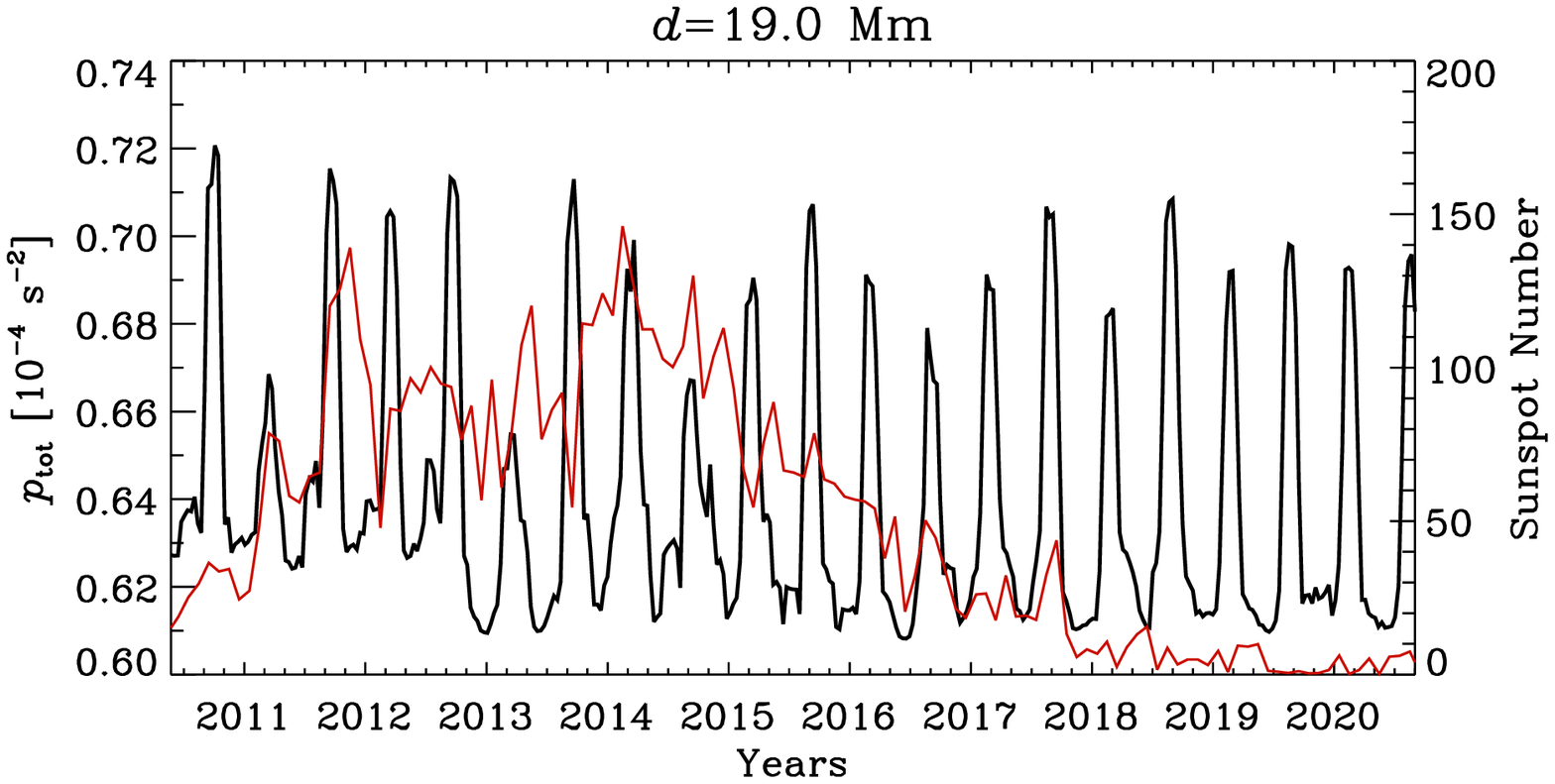}
    \includegraphics[width=0.49\textwidth]{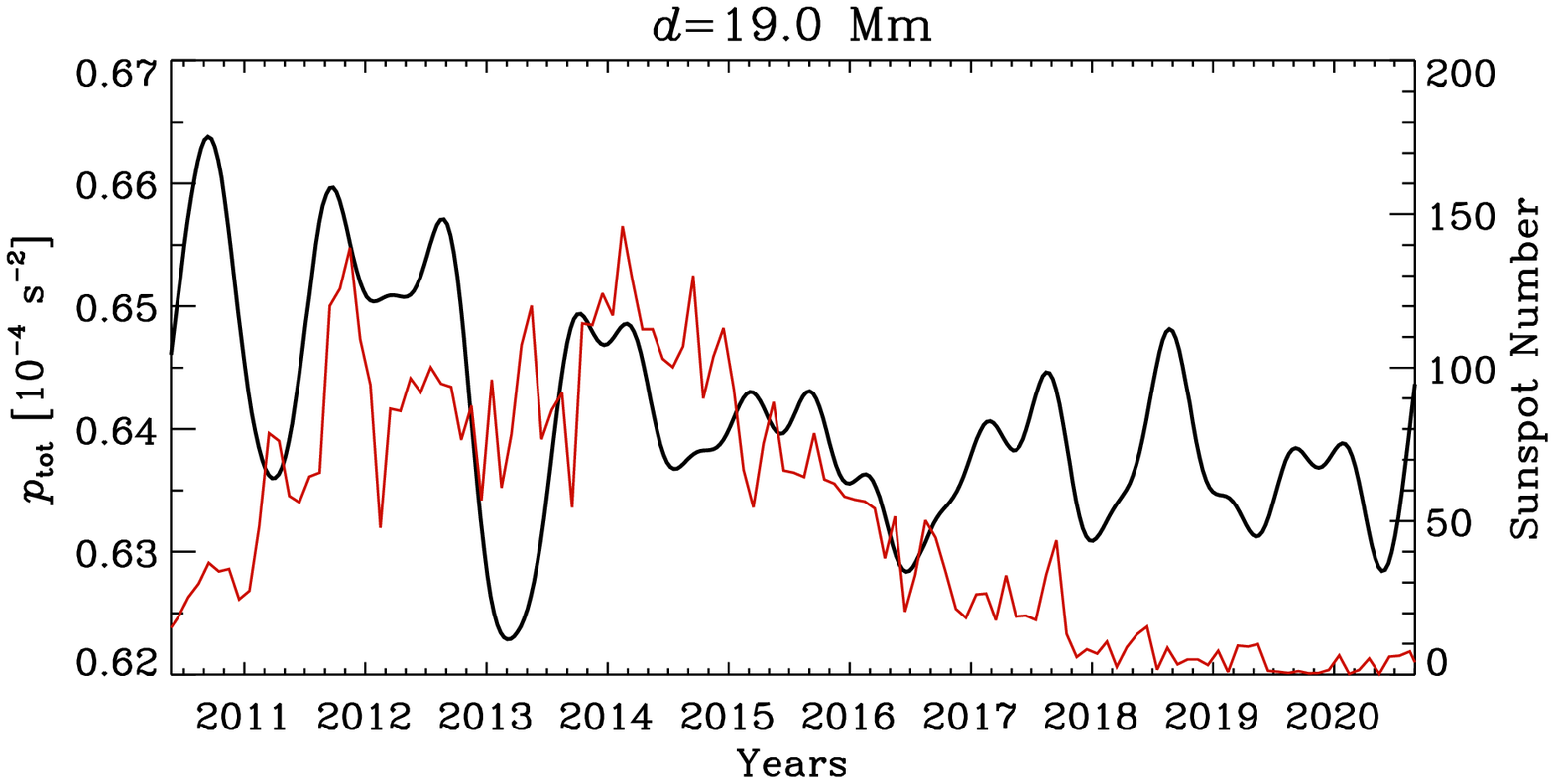}
    \caption{Time variation of the total power of all the harmonics for three depths (indicated at the top of each panel). Left: unfiltered; right: filtered by applying the Butterworth filter with $f_\mathrm H=14, n=4$. The red curve in each panel represents the monthly averaged sunspot number.
	\label{time_var}}
\end{figure*}

According to the \citet{Jeans_1923} formula, the full wavelength of the harmonic $Y_l^{m}$ on a sphere of radius $r$ is
	\begin{equation}
		\lambda=\frac{2\pi r}{\sqrt{l(l+1)}}
		\label{Jeanseq}
	\end{equation}
(the layer that we consider is much thinner than the convection zone, and $r$ can be put equal to the radius of the Sun, $R_\odot$). This wavelength, determined by the degree, $l$, of the spherical harmonic $Y_l^m$, can be used as an estimate of the characteristic size of the flow structures corresponding to this harmonic.
	
\begin{figure} 
	\centering
	\includegraphics[width=0.5\textwidth]{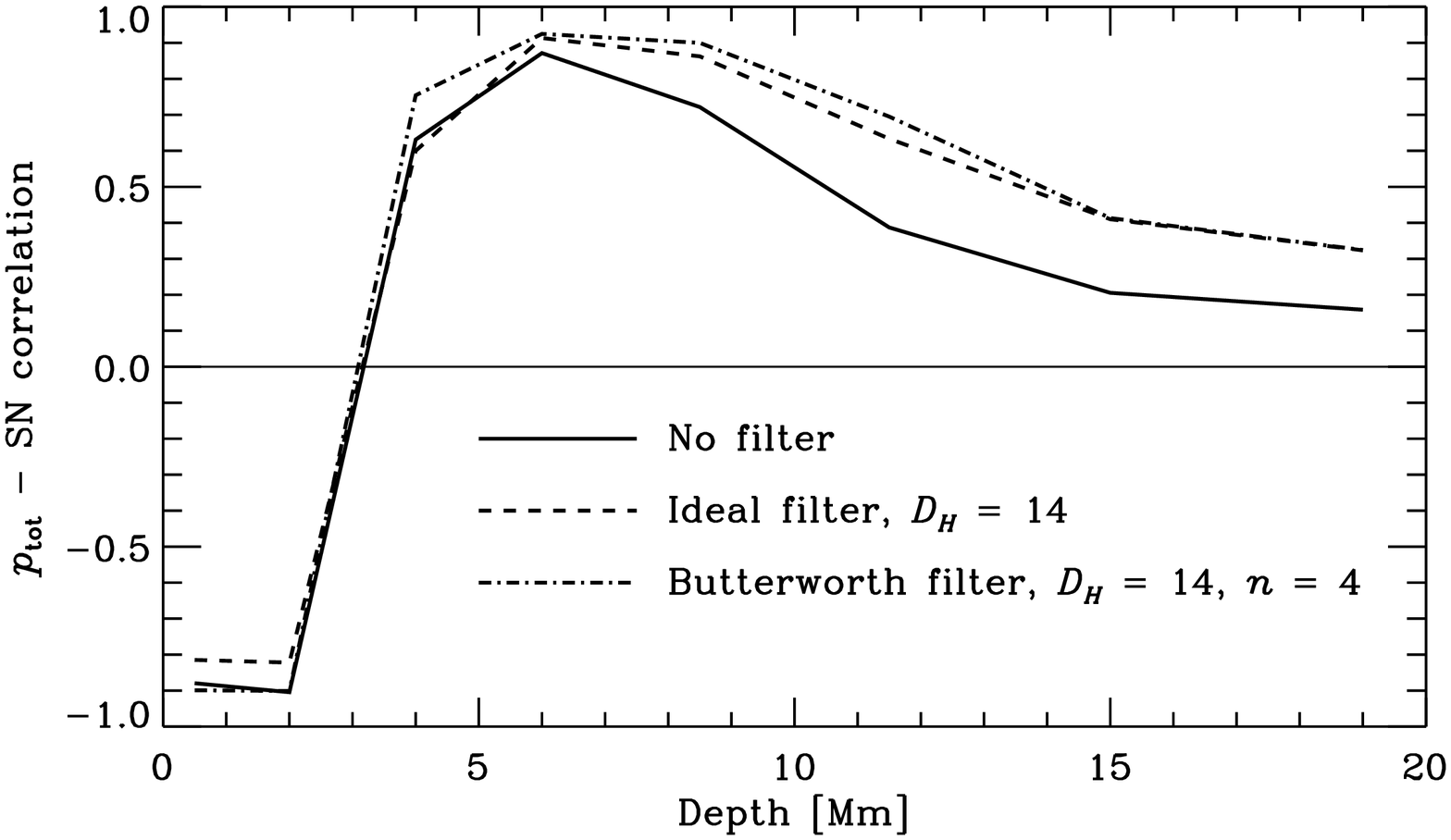}
	\caption{Correlation between the solar-cycle variations of the total power of all the flow harmonics and the monthly averaged sunspot number.\label{correl}}
\end{figure}

	\section{Results}\label{results}
	
	\subsection{Flow Scales}\label{scales}
	
A visual inspection of the divergence maps suggests that the characteristic scale of convection flows increases with depth (Figure~\ref{images}). A typical series of the power spectra of solar convective flows is presented in Figure~\ref{spectra} for different depths. The spectral range of degrees $l$ (or the section of the main spectral peak by an $m=\mathop{\mathrm{const}}$ line) and the corresponding range of scales $\lambda$ are fairly wide in shallow layers. As $d$ increases, the spectral peak narrows and shifts into the low-$l$ (long-wavelength) region. Such behavior can naturally be explained in terms of a superposition of differently scaled flows: the supergranular-scale flows localized in these layers coexist with the upper parts of giant convection cells but do not manifest in deeper levels. In the upper layers, the short-wavelength component of the most energetic harmonics has $l\sim 130$, which, according to Equation~(\ref{Jeanseq}), corresponds to supergranular scales, $\lambda\sim 30$~Mm. In the deeper layers, the largest-scale energetic harmonics correspond to $l\sim 10$. The corresponding wavelength range is broad and centered at 300~Mm, which is a giant-cell scale. These scales are pronounced in the bottom half of the considered layer. They are not so noticeable near the surface because of the more energetic flows of smaller scales; however, the power values for the largest scales in the upper and deep levels are comparable (see below).

Since the harmonics with $l=m$ are sectorial (latitude-independent), the fact that the main spectral peak approaches the $l=m$ line with the  increase of $d$ can be interpreted as a tendency toward the emergence of meridionally elongated, banana-shaped convection structures.
	
According to Parseval's theorem, the integrated power of the flow represented by the spectrum (\ref{series}) is
\begin{equation}
p_\mathrm{tot} \equiv \int\limits_\Omega f^2 \mathrm d\Omega =\sum_{l=0}^\infty \sum_{m=-l}^l |A_{lm}|^2.
\end{equation}
The normalized interior sum
\begin{equation}
p_l=\frac{1}{2l+1}\sum_{m=-l}^l |A_{lm}|^2
\label{powerph}
\end{equation}
{is, by definition, the power per degree $l$ and per steradian \citep[see, e.g., Equations~(B.94) and (B.95) on page 858 in][]{Dahlen_Tromp_1998}. Such a normalization is chosen because it ensures a ``flat'' spectral representation, $p_l=1/4\pi$, of a Dirac delta function on the unit sphere, $(\sin\theta)^{-1}\, \delta(\theta-\theta^\prime)\,\delta(\varphi-\varphi^\prime)$.}
{In addition, we consider the total power of degree $l$,}
\begin{equation}
p^\Sigma_l=\sum_{m=-l}^l |A_{lm}|^2.
\label{powertot}
\end{equation}
{The running averages of $p_l$ and $p^\Sigma_l$ are shown as functions of $l$ in Figure~\ref{m-averages}. The uncertainty of the results is calculated as the standard deviation of the power $p_l$ from its running average and is indicated in Figure~\ref{m-averages} by dotted curves.}

{The helioseismic inferences yield estimates of the flow velocities averaged with kernels of characteristic widths increasing with depth. To evaluate the possible effect of the averaging-kernel broadening with depth, we computed the spectra of a convective-velocity-divergence field for $d=0.5$~Mm smoothed with Gaussian kernels whose widths vary over the same range as do the kernels used in helioseismological inversions \citep{Couvidat_etal_2005}. The $p_l$ spectra corresponding to the extreme kernel-width values, $w=17$ and $38$~Mm, are displayed in the bottom right panel of Figure~\ref{m-averages} and demonstrate only a moderate shift of the spectrum to longer wavelengths with depth, which does not affect our conclusions.}

In the context of detecting the superposition of differently scaled structures, it is worth comparing the power values for the largest-scale harmonics at different depths. The $p_l$ spectra demonstrate the displacement of the main peak to longer wavelengths with depth. The peak of the spectrum for $d=19$~Mm is near $l\approx 13$, its height is about $8.5\times 10^{-9}$~s$^{-2}$.  At $d=0.5$~Mm, the power values for such wavenumber values are about $5\times 10^{-9}$~s$^{-2}$. Similar spectral-power values in this long-wavelength range are also typical at the intermediate depths. Therefore, the flows characterized by the horizontal scales in the range $\lambda\sim 200\,-\,300$~Mm have comparable power values in the whole depth range. However, near the surface, the supergranulation-scale flows are much more powerful. The spectrum thus represents a superposition of flow components with widely differing scales: the supergranular-scale components are localized in the upper layers, while the weaker larger-scale components are extended from deep to shallow layers.
	
	\subsection{Time variation}
	
The total power of the flow, $p_\mathrm{tot}$, measured by the integral of the power spectrum over both $l$ and $m$, exhibits considerable variations in the course of the solar activity cycle. These variations for three $d$ values are shown in the left column of Figure~\ref{time_var} along with the monthly averaged sunspot number. Visually, these two quantities appear to vary nearly in antiphase at the upper level, $d=2$~Mm; nearly in phase at the intermediate level, $d=6$~Mm (where a notable oscillation with a half-year period, 0.063~$\mu$Hz, contaminates the $p_\mathrm{tot}$ variation); and without any definite correlation with the activity level at the bottom level, $d=19$~Mm. The half-year oscillation seems to stem from the variation of the inclination of the Sun's rotational axis to the line of sight. To remove this oscillation and isolate the physically conditioned $p_\mathrm{tot}$ variations on time scales of the order of the activity cycle, we apply a Fourier low-pass-filtering procedure.

Specifically, we calculate the fast Fourier transform of $p_\mathrm{tot}$ as a function of time, then filter the resultant spectrum multiplying it by an appropriate filtering function, $H(f)$ ($f$ being the frequency), and subject the filtered spectrum to the inverse Fourier transform. We use two filters. The first one is an ideal filter,
$$H(f)=\left\{
\begin{aligned}
&1\quad \mathrm{if}\ f \le f_\mathrm H,\\
&0\quad \mathrm{otherwise.}
\end{aligned}
\right.$$
An alternative is a Butterworth low-pass filter
$$H(f)=\frac{1}{1+(f/f_\mathrm H)^{2n}},$$
where $f_\mathrm H$ is a high-wavenumber-cutoff frequency; we assume $f_\mathrm H=0.05$~$\mu$Hz and $n=4$. The algorithms of discrete Fourier transform yield spectra periodic in frequency, the second half of the period corresponding to the negative frequencies decreasing in their absolute magnitude from a maximum (typically put equal to the Nyquist frequency in signal processing) to zero. Accordingly, the filtering function is extended to the second half.
	
The time dependences of $p_\mathrm{tot}$ obtained using the Butterworth filtering are shown in the right column of Figure~\ref{time_var}. We can see that not only becomes the correlation between the sunspot number and the total power more pronounced, especially at the medium depth, $d=6$~Mm, but it emerges even at the lowest level, $d=19$~Mm, where it was not notable without filtering. To quantify the possible solar-activity dependence of the convection-flow energy, we calculate the coefficient of correlation between the two above-mentioned quantities for each depth. As can be seen from Figure~\ref{correl}, both the anticorrelation at $d=2$~Mm and positive correlation at $d=6$~Mm are fairly high and do not significantly depend on the choice of the filter. The two temporal variations are best correlated if the Butterworth filter is used---in this case, the correlation coefficient is $-0.901$ at $d=2$~Mm, 0.925 at $d=6$~Mm, 0.695 at $d=11.5$~Mm, and 0.324 at $d=19$~Mm.
	
    \section{Conclusion and Discussion}
	
We have analyzed the spatial spectra of the horizontal-velocity-divergence field at depths in the solar convection zone ranging from 0 to 19~Mm. The range of flow scales is fairly wide in shallow layers. As the depth increases, this range narrows, and the main peak shifts into the long-wavelength region. While the shortest-length scales of the most energetic harmonics in the upper layers correspond to supergranular scales, $\sim 30$~Mm, the largest scales in the deepest layers are about 300~Mm, which is definitely a giant-cell scale. The large-scale components are not clearly noticeable in the top layers because of the strong supergranulation component, but their power is of the same order of magnitude as in the deep layers. Such behavior can naturally be interpreted in terms of a superposition of differently scaled flows localized in different depth intervals within the convection zone. In addition, there is some tendency toward the emergence of meridionally elongated convection structures in the deeper layers.

We have also considered the time variation of the integrated spectral power of the flow at different levels with the solar activity cycle. To remove the half-year oscillation with a frequency of 0.063~$\mu$Hz, attributed to the variations in the inclination of the solar rotational axis to the line of sight, we applied a spectral filtering procedure with a Butterworth low-pass filter. The results show that the time variation of the total power is anticorrelated with the sunspot number in the shallow layers and positively correlated at larger depths.

The detected relationship between the solar activity and convective-velocity power can likely be interpreted in terms of magnetic-field-induced differences in the flow structure at different depths during high- and low-activity periods. This important issue calls for further investigation.
	
    \section*{acknowledgments}
We are grateful to G.~Guerrero and A.M.~Stejko for providing the data of their numerical simulations. The helioseismological data used here were derived from HMI observational data available courtesy of the NASA/SDO and HMI science teams. We also used sunspot-number data from the World Data Center for Sunspot Index, and Long-term Solar Observations (WDC-SILSO), Royal Observatory of Belgium, Brussels. The work partially supported by NASA grants NNX14AB70G, 80NSSC20K1320, 80NSSC20K0602.

	\bibliographystyle{aasjournal}
	\bibliography{Scales_1}

\end{document}